\documentclass[sigconf]{acmart}

\usepackage{latexsym}
\usepackage{graphicx}
\graphicspath{{./images/}}
\usepackage{booktabs} % for formal tables
\usepackage{color}  % for coloring text
\usepackage{amsmath}  % for aligning equations
\usepackage{subcaption}
\usepackage{caption}
\usepackage{tikz}
\usepackage{colortbl} % for color in tables
\usepackage{framed}
\usepackage{multirow}
\usepackage{multicol}
\usepackage{hyperref}
\usepackage{url}
\usepackage{balance}
\usepackage{verbatim}
\usepackage{cancel}
\usepackage{xspace} % for correcting space after macro commands
\usepackage[ruled,linesnumbered,vlined]{algorithm2e}
\usepackage{bbold} % for writing mathbb{1}

\newcommand{\struct}[1]{\texttt{\small #1}}
\newcommand{\utterance}[1]{\textit{#1}}
\newcommand{\phrase}[1]{\textit{``#1''}}

\newcommand{\myparagraph}[1]{\noindent \textbf{#1}.}

\newenvironment{Snugshade}[1][236,236,236]{
	\setlength{\itemsep}{0pt}
	\setlength{\parsep}{0pt}
	\setlength{\topsep}{0pt}
	\setlength{\partopsep}{0pt}
	\setlength{\leftmargin}{1.5em}
	\setlength{\labelwidth}{0em}
	\setlength{\labelsep}{0em} 
	\setlength{\parskip}{0pt}
	\definecolor{shadecolor}{RGB}{#1}%
	\begin{snugshade}
	}{%
	\end{snugshade}%
}

\newcommand{\conquer}{\textsc{Conquer}\xspace}
\newcommand{\convquestionsref}{\textsc{ConvRef}\xspace}

\newcommand{\convex}{\textsc{Convex}\xspace}
\newcommand{\convquestions}{\textsc{ConvQuestions}\xspace}
\newcommand{\elq}{\textsc{Elq}\xspace}

\copyrightyear{2021}
\acmYear{2021}
\acmConference[SIGIR '21]{Proceedings of the 44th International ACM SIGIR Conference on Research and Development in Information Retrieval}{July 11--15, 2021}{Virtual Event, Canada}
\acmBooktitle{Proceedings of the 44th International ACM SIGIR Conference on Research and Development in Information Retrieval (SIGIR '21), July 11--15, 2021, Virtual Event, Canada}
\begin{document}
	
	% \title{Leveraging Implicit Feedback via Reformulations in Conversational Question Answering over Knowledge Graphs} % with Reinforcement Learning}
	% \title{ConQueR: An RL-Framework for Leveraging Reformulations in Conversational Question Answering over Knowledge Graphs} % with Reinforcement Learning}
	% \title{ConQueR: An RL Model for Leveraging Reformulations in Conversational Question Answering over Knowledge Graphs} % with Reinforcement Learning}
	% \title{Leveraging Reformulations with Reinforcement Learning for Conversational Question Answering over Knowledge Graphs}
	% \title{ConQueR: Using Reformulations with Reinforcement Learning for Conversational Question Answering over Knowledge Graphs}
	\title{Reinforcement Learning from Reformulations in\\
		Conversational Question Answering over Knowledge Graphs}
	
	\author{Magdalena Kaiser}
	%\authornote{Dr.~Trovato insisted his name be first.}
	\affiliation{%
		\institution{Max Planck Institute for Informatics, Germany}
		% \\ Saarland Informatics Campus, Germany}
	}
	\email{mkaiser@mpi-inf.mpg.de}
	
	\author{Rishiraj Saha Roy}
	\affiliation{%
		\institution{Max Planck Institute for Informatics, Germany}
		% \\ Saarland Informatics Campus, Germany}
		%\streetaddress{Address}
		%\city{City}
		%\state{Country}
	}
	\email{rishiraj@mpi-inf.mpg.de}
	
	\author{Gerhard Weikum}
	\affiliation{%
		\institution{Max Planck Institute for Informatics, Germany}
		% \\ Saarland Informatics Campus, Germany}
	}
	\email{weikum@mpi-inf.mpg.de}
	
	{\tiny }\renewcommand{\shortauthors}{Kaiser et al.}
	
	\newcommand\BibTeX{B{\sc ib}\TeX}
	
	\newcommand{\squishlist}{
		\begin{list}{$\bullet$}
			{ \setlength{\itemsep}{0pt}
				\setlength{\parsep}{3pt}
				\setlength{\topsep}{3pt}
				\setlength{\partopsep}{0pt}
				\setlength{\leftmargin}{1.5em}
				\setlength{\labelwidth}{1em}
				\setlength{\labelsep}{0.5em} } }
		
		\newcommand{\squishend}{
	\end{list}  }
	
	% !TEX root = ../2021-sigir-fp-conquer.tex
\begin{abstract}
	
The rise of personal assistants has made conversational question answering (ConvQA) a very popular mechanism for user-system interaction.
State-of-the-art methods for ConvQA over knowledge graphs (KGs) can only learn from crisp question-answer pairs found in popular benchmarks. In reality, however, such training data is hard to come by: users would rarely mark answers explicitly as correct or wrong. In this work, we take a step towards a more natural learning paradigm -- from noisy and implicit feedback via question reformulations. A reformulation is likely to be triggered by an incorrect system response, whereas a new follow-up question could be a positive signal on the previous turn's answer. We present a reinforcement learning model, termed \conquer, that can learn
% exclusively
from a conversational stream of questions and reformulations. \conquer models the answering process as multiple agents walking in parallel on the KG, where the walks are determined by actions sampled using a policy network. This policy network takes the question along with the conversational context as inputs and is trained via noisy rewards obtained from the reformulation likelihood. To evaluate \conquer, we create and release \convquestionsref, a benchmark with about $11k$ natural conversations containing around $205k$ reformulations. Experiments show that \conquer successfully learns to answer conversational questions from noisy reward signals, significantly improving over a state-of-the-art baseline.		
\end{abstract}


	
	% \keywords{Question Answering, Knowledge Graphs, Conversations, \\ User Feedback}
	
%	\fancyhead{}
	
	\maketitle
	
	%% !TEX root = ../2021-sigir-fp-conquer.tex
\section{Introduction}
\label{sec:intro}

% \subsection{Background and motivation}
% \label{subsec:background}

\begin{figure} [t]
	\centering
	\includegraphics[width=\columnwidth]{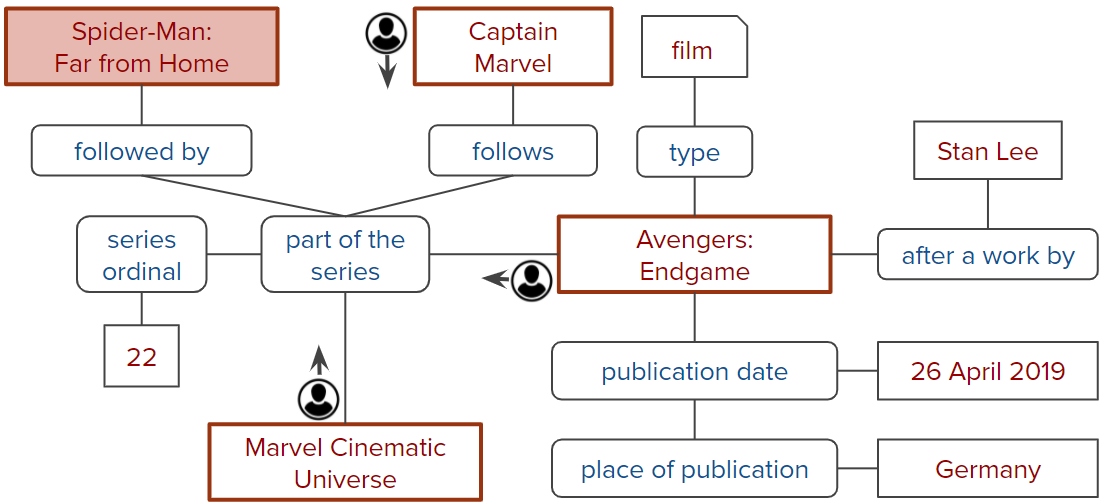}
% 	\caption{An illustration of the KG snippet from Wikidata required for answering $q_1$ and $q_2$. Possible agents for $q_{22}$ (\utterance{What came next in the series?}) at various nodes are shown, with possible walking directions for each agent. The colored box (\struct{Spider-man: Far from Home}) shows the correct answer.}
	\caption{KG excerpt from Wikidata required for answering $q_1$ and $q_2$. Agents for $q_{22}$ 
	(\utterance{What came next in the series?}) % at various nodes are shown, with possible walking directions for each agent. 
	are shown with possible walk directions.
	The
	% white
	colored box (\struct{Spider-man: Far from Home}) is the correct answer.}
	\label{fig:kg}
	\vspace*{-0.65cm}
\end{figure}

\myparagraph{Background and motivation} Conversational question answering (ConvQA)
has become a convenient and natural
% is a natural
mechanism of satisfying information needs
that are too complex or exploratory to be formulated in
a single shot~\cite{choi2018quac,saha2018complex,reddy2019coqa,guo2018dialog,qu2020open,kaiser2020conversational}.
ConvQA operates in a multi-turn, sequential mode of information access:
utterances in each turn are ad hoc and often incomplete,
with implicit context that needs to be inferred from prior turns.
% in follow-up questions~\cite{christmann2019look}.
%This simulates a human-like communication experience with personal 
%assistants
%like Cortana, Alexa, Siri, or the Google Assistant.
When the information needs are 
%factoid 
%or objective in nature
fact-centric 
(e.g., about cast of movies, clubs of soccer players, etc.),
% sequels of books, etc.),
%the best underlying answering resource to go for would
a suitable data source to retrieve answers from 
are large knowledge graphs (KG) such as Wikidata~\cite{vrandevcic2014wikidata},
Freebase~\cite{bollacker2008freebase},
%be large curated knowledge graphs (KG) like Wikidata~\cite{vrandevcic2014wikidata},
DBpedia~\cite{auer2007dbpedia}, or YAGO~\cite{suchanek2007yago}.
% , or %or industrial KGs,
Fig.~\ref{fig:kg} shows a small excerpt of the Wikidata
KG using a simplified graph representation, with red nodes for entities and blue nodes for
relations.
%which capture entities (red nodes in Fig.~\ref{fig:kg}) and 
%%organize information as 
%relationships (blue nodes).
% in Fig.~\ref{fig:kg}).
%between entities (red nodes).
This paper addresses
%conversational QA over KGs (ConvKGQA) 
ConvQA over KGs,
%are the focus of 
%this work, 
where system responses are usually entities. 

\noindent{\bf Example.}
%An example of 
An ideal conversation with five \textit{turns}
%is shown below.
could be
as follows ($q_i$ and $ans_i$ are questions and answers at turn $i$, respectively):
\begin{Snugshade}
	\noindent	$q_{1}$: \utterance{When was Avengers: Endgame released in Germany?}\\
	$ans_1$: \textit{24 April 2019}\\
	\noindent	\textbf{$q_{2}:$} \utterance{What was the 
		next from Marvel?}	 \\	
	\noindent	\textbf{$ans_2$:} \textit{Spider-Man: Far from Home} 	\\
	\noindent \textbf{$q_{3}$ :} \utterance{Released on?}\\
	\noindent 	\textbf{$ans_3$:} \textit{04 July 2019} \\
	\noindent	\textbf{$q_{4}$:} \utterance{So who was Spidey?}	 \\	
	\noindent	\textbf{$ans_4$:} \textit{Tom Holland} 	\\
	\noindent	\textbf{$q_{5}:$} \utterance{And his girlfriend was played by?}	 \\	
	\noindent	\textbf{$ans_5$:} \textit{Zendaya Coleman}
\end{Snugshade}
%GW: this is state of the art, interrupt the flow here
%Research on ConvKGQA is still in its
%infancy~\cite{saha2018complex,guo2018dialog,christmann2019look,shen2019multi}:
%it is not surprising that even state-of-the-art systems	
%%are prone to substantial errors, even on simple
% questions.
% (an exploration of the live prototype of the \convex system
% at \url{https://convex.mpi-inf.mpg.de/} would convince the reader).
%This is due to the inherent challenges posed by the setup where
%he bulk of the 
%As shown in this conversation, 
Utterances can be
%could be
colloquial ($q_4$)
% spidey
and incomplete ($q_2, q_3$),
% next what, Marvel what
and inferring the proper context is a challenge ($q_5$). 
%whose girlfriend? Tom's? Spiderman's?
% probably the reader is not yet ready for the part below
%Concretely, many near misses are caused by misleading semantic matches
%in KG neighborhoods established
%by the conversational context (see Fig.~\ref{fig:kg} for an illustration).
%For instance, \textit{next} in $q_2$ could easily match \textit{after a work by} in the KG,
%and return \textit{Stan Lee} as the answer.
%
%In such a scenario, 
Users can provide \textit{feedback} in %a natural way in 
the form of
\textit{question reformulations}~\cite{ponnusamy2020feedback}: 
%in case of
%an unsatisfactory or incorrect answer,
%a user would
%issue a reformulation to the system, 
when an answer is incorrect, users may rephrase the question,
hoping for better results.
While users never know the correct answer upfront,
they may often guess non-relevance when the answer does not match
the expected type (director instead of movie) or from 
additional background knowledge.
% (a user may ask about
%a 2019 Avengers movie, yet receive information about one that she knows is from 2012).
%So in reality, 
So, in reality, turn 2 in the conversation above could 
%look like
%(intent below simply refers to a new information need):
become expanded into:
% follows:
\begin{Snugshade}
	\noindent	\textbf{$q_{21}:$} \utterance{What was the next from Marvel?} \textbf{(New intent)}	 \\	
	\noindent	\textbf{$ans_{21}$:} \textit{Stan Lee}  \textbf{(Wrong answer)}\\	
	\noindent   \textbf{$q_{22}$ :} \utterance{What came next in the series? } \textbf{(Reformulation)}\\		
	\noindent 	\textbf{$ans_{22}$:} \textit{Marvel Cinematic Universe}  \textbf{(Wrong answer)}\\		
	\noindent 	\textbf{$q_{23}$:} \utterance{The following movie in the Marvel series? }  \textbf{(Reformulation) }\\	
	\noindent 	\textbf{$ans_{23}$:} \textit{Spider-Man: Far from Home}  \textbf{(Correct answer)}\\	
	\noindent 	\textbf{$q_{31}$:} \utterance{Released on?}	  \textbf{(New intent)}	
\end{Snugshade} 
%Without a traditional Web interface where clicks or
%	mouse hovers could be recorded as additional signals
%	of implicit feedback, appropriate harnessing of reformulations
%	is
%of utmost importance in personal assistants in mobiles or home devices.
%
% no time for this in this submission
% Such reformulations are also
% key to understanding idiosyncratic behavior of \textit{individual users}.
% Since the typical user hardly provides explicit feedback when
% answers are correct, reliance on these
% implicit cues are even more worthwhile.
% This motivates our current contribution.

%Note that
%the conversational setting is \textit{more general} than the usual KG-QA setup
%dealing
%with well-formed questions, and thus the proposed contribution in
%the conversational setting can be applied as-is to the standard paradigm.

% \subsection{Limitations of state-of-the-art}
% \label{subsec:limits}

\myparagraph{Limitations of state-of-the-art} Research on ConvQA over KGs is still in its
infancy~\cite{saha2018complex,guo2018dialog,christmann2019look,shen2019multi} --
in particular, there is virtually no work on considering user signals when intermediate utterances lead to unsatisfactory responses as indicated in the above conversation with reformulations.
A few works on QA over KGs has exploited user interactions for online learning
% \cite{abujabal2018never,kratzwald2019learning,zhang2019interactive},
\cite{abujabal2018never,zhang2019interactive},
but this is limited to confirming the correctness of answers which can then
augment the training data of question-answer pairs.
%
%\textbf{User feedback.} The lack of large volumes of real 
%usage data for ConvQA over KGs makes on-the-job learning from user feedback a 
%critical feature, that is conspicuously missing in 
%state-of-the-art systems. A small number of QA systems over KGs and text 
%leverage user feedback~\cite{abujabal2018never,kratzwald2019learning,zhang2019interactive}.
%But this is limited to 
%explicit positive 
%feedback on the answer, that directly results in augmented training sets
%for a supervised learning setup. This is mainly driven by the long-standing 
%idea of learning only from question-answer pairs in QA 
%benchmarks~\cite{berant2013semantic,christmann2019look,rajpurkar2016squad,usbeck20189th}.
%In reality though, such datasets would be hard to curate: the average Web user
%rarely provides explicit relevance labels, making the exploitation of 
%implicit cues the more practical approach.
%
%MK: add amazon feedback paper somewhere here?
Reformulations
% of queries,
as an implicit feedback signal have been
leveraged for web search queries~\cite{joachims2007evaluating,ren2018conversational},
but that setting is very different from QA over KGs.
IR methods rely on observing clicks (and their absence) on ranked lists of documents. 
This
%assumption
does not carry over to
% ConvQA over KGs.
typical QA tasks -- especially over voice interfaces
with single-entity responses at each turn and no
explicitly positive click-like signal.

% \subsection{Approach and contributions}
% \label{subsec:approach}

\myparagraph{Approach}
% \textbf{\conquer}. 
%We now summarize our approach, 
We present
\conquer
(\underline{Con}versational \underline{Que}stion answering with \underline{R}eformulations),
a new method for learning from implicit user feedback in
ConvQA over KGs.
% The method is 
% termed \conquer, that stands for \underline{\textbf{Con}}versational 
% \underline{\textbf{Que}}stion answering with
%	\underline{\textbf{R}}eformulations.	
\conquer is based on reinforcement learning (RL) and is designed to continuously learn from question reformulations as a cue that the previous system response was unsatisfying.
RL methods have been pursued for KG reasoning and multi-hop QA over KGs~\cite{das2018go,lin2018multi,qiu2020stepwise}. 
However, {\em conversational} QA is very different from
% the
% setup in
% % system model of
% these prior works. 
these prior setups.

%%%GW: now dive into more technical depth
Given the current (say $q_{22}$) and the previous utterances ($q_1, q_{21}$),
% conversational utterance , and the previous history 
\conquer creates and maintains a
set
% scored list
% priority queue
% it is perhaps okay to use priority as entities are previous turns are de-prioritized, and new ones are enqueued 
of \textit{context entities} from the KG that are the most relevant to the conversation so far.
% While the current question represents the lexical context, these nodes represent the KG-context for the conversation, and together they should suffice for determining the answer.	
It then positions \textit{RL agents} at each of these 
% \textit{start points},
context entities,
that simultaneously walk over the KG to other entities in their respective neighborhoods. End points of these walks are candidate answers for this turn and are aggregated for producing the final response. 
Walking directions (see arrows in Fig.~\ref{fig:kg}) are decided by sampling actions from a policy network that takes as input
%BERT
i) encodings of
% the current utterance,
utterances,
and ii) KG facts involving the
% start points
context entities.
% and the history
%.
% Wherever token sequence matters (current utterance, conversational history) we use LSTM encodings.
The policy network is trained via noisy rewards obtained from reformulation likelihoods estimated by a fine-tuned BERT predictor.
Experiments on our \convquestionsref benchmark, that we created from conversations between a system and real users, demonstrate the viability of our proposed learning method \conquer and its superiority over
a state-of-the-art baseline.
% \convex~\cite{christmann2019look}. 
%The benchmark and
%results,
%a demo
%, and the link to our code on GitHub 
%are accessible at \url{https://conquer.mpi-inf.mpg.de}.
%The code is available at \url{https://github.com/magkai/CONQUER}.
The benchmark and demo are at: https://conquer.mpi-inf.mpg.de and the code is at: https://github.com/magkai/CONQUER.
%Benchmark and
%results,
%demo
%, and the link to our code on GitHub 
%available at: \url{https://conquer.mpi-inf.mpg.de}, the code at: \url{https://github.com/magkai/CONQUER}.

\textbf{Contributions.} 
%We make the following salient contributions:
Salient contributions of this work are:
\squishlist
\item A question answering method that can learn from a conversational stream in the \textit{absence of gold answers}; 
\item A \textit{reinforcement learning} model for QA with rewards based on implicit feedback in the form of question \textit{reformulations;}
%MK: demo needs to be mentioned as well (maybe not as contribution)+future conquer website
\item A \textit{reformulation detector} based on BERT that can classify a follow-up utterance as a reformulation or new intent;
\item A new \textit{benchmark collection with reformulations}
for ConvQA over KGs, comprising
about $11k$ conversations with more than $200k$ turns in total, out of which $205k$ are reformulations.
% \item A reformulation predictor based on a fine-tuned BERT model.
%MK: TODO: include github+future conquer website + demo needs to be mentioned as well (probably not as contribution)
\squishend

	\section{Model and Architecture}
\label{sec:setup}

\subsection{\conquer KG representation}
\label{subsec:kg-model}

\myparagraph{Basic model} A knowledge graph (KG) $K$ is typically stored as an RDF database organized into $\langle S, P, O \rangle$ (subject, predicate, object) triples (facts), where $S$ is an entity (e.g., \struct{Avengers: Endgame, Stan Lee}), $P$ is a predicate (e.g., \struct{part of series, publication date}), and $O$ is an entity, a type (e.g., \struct{film, country}) or a literal (e.g., \struct{26 April 2019, 22}). 
% For a graph-based representation see Fig.~\ref{fig:kg}.
% Each KG item (entities, predicates, types, literals) is made a node and there are edges between items to capture the original triples.
In \conquer, we wish to leverage the entire KG for answering. For that we need to go beyond triples and consider \textit{$n$-ary facts}, as discussed below.

\myparagraph{Qualifier model} Large KGs like Wikidata also contain $n$-ary facts that involve more than two entities. Examples are: cast information involving a movie, a character role and a cast member or movie trilogy information requiring the movies, their ordinal numbers and the name of the series.
% or, award information including a person, an award, and a movie.
$n$-ary facts are typically represented as a \textit{main fact} enhanced with \textit{qualifiers}, that are (possibly multiple) auxiliary $\langle$predicate, object$\rangle$ pairs adding contextual information to the main fact. For example, in Fig.~\ref{fig:kg}, the path in the graph connecting \struct{Avengers: Endgame} to \struct{part of the series} and over to \struct{Marvel Cinematic Universe} represent a main fact, which is contextualized by the path connecting this main fact to the qualifier predicate node \struct{followed by} and on to the qualifier object node \struct{Spider-man: Far from Home}. The path with \struct{series ordinal} and \struct{22} is another qualifier for the same main fact. The KG schema determines which part of the fact is considered main and which part a qualifier. This is often interchangeable, and in practice each provides context for the other. A large part of QA research disregards qualifiers, but they contain valuable information~\cite{nguyen2014dont,oguz2020unified,leeka2016quark,hernandez2015reifying,galkin2020message} and constitute a substantial fraction of Wikidata and other KGs. Without considering qualifiers in Wikidata, we would not be able to answer \textit{any of the questions from $q_1$ through $q_5$}.

\begin{figure} [t]
	\centering
	\includegraphics[width=\columnwidth]{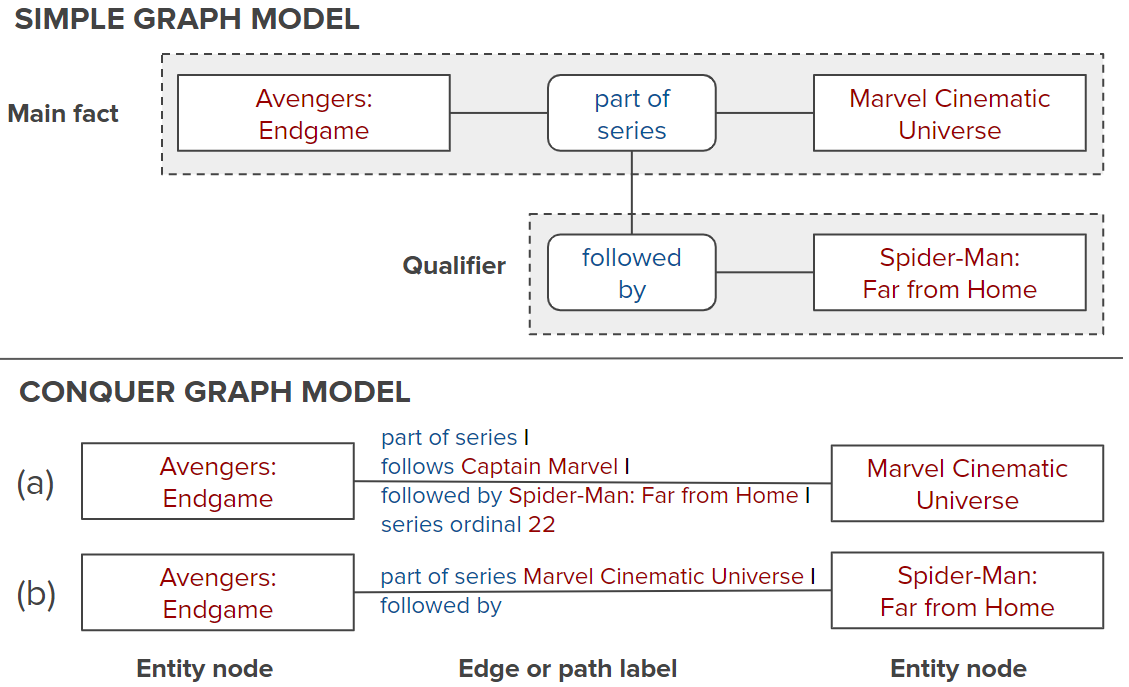}
	% \vspace*{-0.7cm}
	\caption{\conquer KG representation for using qualifiers.}
	\label{fig:qualifier}
	\vspace*{-0.3cm}
\end{figure}

In the graph representation depicted in Fig.~\ref{fig:kg}, qualifier predicates are directly connected to their main-fact predicates. However, in this representation, an agent walking from entity to entity (e.g., \struct{Avengers} to \struct{Spider-Man}) misses the context of \struct{Marvel} and would be useless for answering $q_2$. In this case, the main-fact triple (\struct{<Avengers: Endgame, part of series, Marvel Cinematic Universe>}) provides necessary context for making sense of the qualifier (\struct{<followed by, Spider-Man: Far from Home>}). To take $n$-ary facts into account during walks by agents, \conquer creates a modified KG representation where entities are nodes and edges between entities are labeled either by connecting predicates (when a fact has no qualifiers, like \struct{<Avengers: Endgame, after a work by, Stan Lee>}) or by \textit{augmented labels} in cases of facts with qualifiers. The latter scenario is visualized in Fig.~\ref{fig:qualifier}. The edge between the main-fact subject \struct{Avengers: Endgame} and the main-fact object \struct{Marvel Cinematic Universe} is augmented by its qualifier information in Fig.~\ref{fig:qualifier} (a). Information from the main fact is also used to augment the connections between the main-fact subject (or object) and qualifier objects, as in Fig.~\ref{fig:qualifier} (b). Connections between qualifier objects are analogously augmented by the main fact. These edge labels or \textit{paths} subsequently become \textit{actions} to be chosen by RL agents. The \conquer graph model is \textit{bidirectional}. %, so agents can walk both ways.

\subsection{ConvQA concepts}
\label{subsec:concepts}

\begin{table} [t]
	\centering
	\resizebox*{\columnwidth}{!}{
		\begin{tabular}{l l}
			\toprule
			\textbf{Notation}				& \textbf{Concept}										                            \\ \toprule
			% $K, E, P, T, L$				& Knowledge graph, entity, predicate, type, literal	                                \\
			$K$				                & Knowledge graph	                                                                \\
			$q, ans$					    & Question and answer 									                            \\			
			% $E, P, T, L$					& Knowledge graph, entity, predicate, type, literal		                            \\			
			% $S, P, O$						& Subject, predicate, object							                            \\
			% $\mathcal{K}, \mathcal{P}$	& Processed KG, processed predicate set				                                \\
			$C$								& Conversation											                            \\
			$I$								& Intent												                            \\
			$q_{j1}$	                    & First question in intent $j$							                            \\
			$\langle q_{jk} | k > 1\rangle$	& Sequence of reformulations in intent $j$				                            \\
			$t$								& Turn 													                            \\ \midrule 
			$q_t^{cxt} \in Q_t^{cxt}$       & Context questions at turn $t$                                                     \\ 
			$e_t^{cxt} \in E_t^{cxt}$       & Context entities at turn $t$                                                      \\ 
			$h_{(\cdot)}$					& Hyperparameters for context entity selection			                            \\		
			$\boldsymbol{q_t}, \boldsymbol{q}^{cxt}_t, \boldsymbol{e}^{cxt}_t$   & Embedding vectors of $q_t, q^{cxt}_t, e^{cxt}_t$ \\ \midrule
		    $s \in \mathcal{S}$             & RL states                                                                         \\
		    $a \in A_{s}$                   & Actions at state $s$                                                              \\
            $\boldsymbol{a}, \boldsymbol{A_s}$   & Embedding vector of $a$, and matrix of all actions at $s$                    \\
			$p \in \mathcal{P}$             & Path labels in $K$                                                                \\
		    $R$                             & Reward                                                                            \\
		    $\boldsymbol{\theta}$           & Parameters of policy network                                                      \\
		    $\pi_{\boldsymbol{\theta}}$     & Policy parameterized by $\boldsymbol{\theta}$                                     \\
            $J({\boldsymbol{\theta}})$      & Expected reward with $\boldsymbol{\theta}$                                        \\		    
%		    $e_{start}, e_{end}$            & Start/End of agent's walk                                                         \\
% 			$sp_t$							& Start points at turn $t$								                            \\ 			
% 			$ep_t$							& End points at turn $t$								                            \\	
			$\boldsymbol{W_1}, \boldsymbol{W_2}$    & Weight matrices in policy network			                                \\
			$\alpha$                        & Step size in REINFORCE update                                                     \\
			$H_\pi(\cdot, s)$   & Entropy regularization term in REINFORCE update                                   \\    
			 $\beta$                & Weight for entropy regularization term \\
			$e^{ans}$                       & Candidate answer entity                                                    \\ 	\bottomrule
	\end{tabular}}
	\caption{Notation for salient concepts in \conquer.}
	\label{tab:notation}
	\vspace*{-1cm}
\end{table}

We now 
%formally
define 
%some of the 
key concepts 
%that will be used throughout
%the rest of this paper. 
for ConvQA below. A notation overview is in Table~\ref{tab:notation} (some concepts are introduced only in later sections).
%ready reference for 
%the useful concepts.

\myparagraph{Question} A question $q$ (aka utterance)
is composed of a sequence of
% $n$
words
% $\langle w_{j} | j=1 \ldots n \rangle$
that is issued by the user to
instantiate a specific 
%intent 
information need
in the conversation.
We make no assumptions on the grammatical correctness of $q$.
Questions may express new information needs or reformulate existing ones.

\myparagraph{Answer} An answer $ans$ is a single or
a (typically small) set of entities (or literals)
% $\{E\}$
from $K$, that the system returns to the user
in response to her question $q$.

\myparagraph{Conversation} A conversation $C$ is a sequence of
questions $\langle q \rangle$ and 
corresponding answers $\langle ans \rangle$. 
$C$ can be perceived as being
organized into a sequence of user intents.

\myparagraph{Intent} Each distinct information need in a conversation $C$
is referred to as an intent $I$. The ideal conversation in 
Sec.~\ref{sec:intro} has five intents $\langle I_1, \ldots, I_5 \rangle$.
Intents are latent and expressed by questions $q$.

\myparagraph{Reformulation}
For a specific intent $I_j$, a user issues reformulation questions
$\langle q_{jk} | k > 1 \rangle$ ($q_{j2}, q_{j3}, \ldots$)
when the system response $ans$ to the first question $q_j$ (equivalently $q_{j1}$) was wrong.
All intents, including the first one, can be potentially reformulated.

\myparagraph{Turn} Each question in $C$, including its reformulations
and corresponding answers, constitutes a turn $t_i$. For instance,
for conversation
$C = \langle q_{11}, ans_{11}, q_{12}, ans_{12}, q_{21}, ans_{21}, q_{22}, ans_{22}, q_{31}, ans_{31} \rangle$,
we have three intents ($I_1, I_2, I_3$), five questions $q_{(\cdot)}$, five answers $ans_{(\cdot)}$, two reformulations ($q_{12}, q_{22}$),
and \textit{five} turns ($t_1, \ldots, t_5$). Thus, $C$ may also be written as $\langle q_{t_1}, ans_{t_1}, q_{t_2}, ans_{t_2}, q_{t_3}, ans_{t_3}, \ldots, q_{t_5}, ans_{t_5} \rangle$. To simplify notation, when we refer to a question at turn $t_i$, we will only use $q_i$ instead of $q_{t_i}$ (analogously for context).

\myparagraph{Context questions} At any given turn $t$, context questions $Q_t^{cxt}$ are the ones most relevant to the conversation so far. 
This set may be comprised of a few of the immediately preceding questions $q_{t-1}, q_{t-2}, \ldots$, or include the first question ($q_1$) in $C$ as well. 

\myparagraph{Context entities} At any given turn $t$, context entities $E_t^{cxt}$ are the ones most relevant to the conversation so far. These are identified using various cues from the question and the KG and form the \textit{start points} of the walks by the RL agents.

\subsection{System overview}
\label{subsec:overview}

\begin{figure} [t]
	\centering
	\includegraphics[width=0.9\columnwidth]{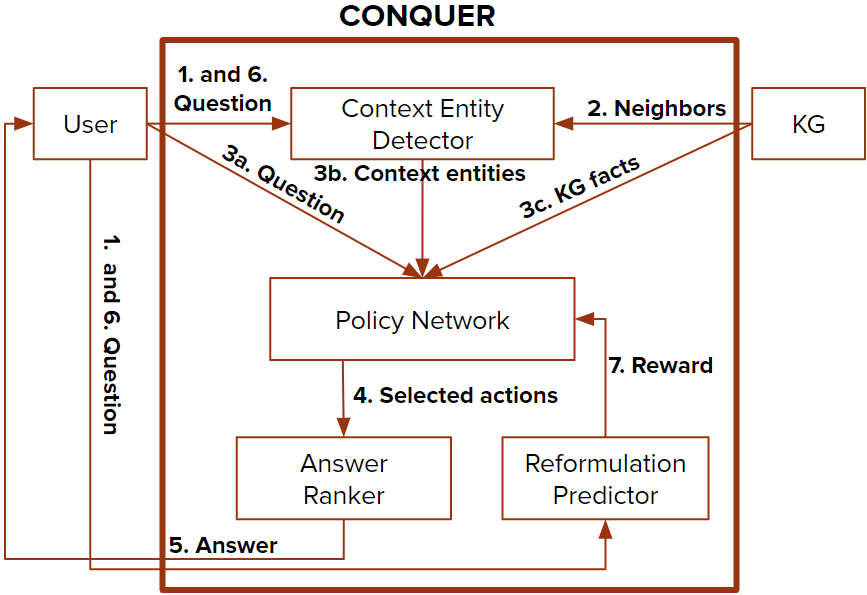}
	% \vspace*{-0.5cm}
	\caption{Overview of \conquer, with numbered steps tracing the workflow. Looping through steps 1-7 creates a continuous learning model. Step 6 denotes a follow-up question.}
	\label{fig:overview}
	\vspace*{-0.6cm}
\end{figure}

The workflow of \conquer is illustrated in Fig.~\ref{fig:overview}. 
% At training time,
First, % At each turn of the conversation, the
% {\em start points} (i.e.,
context entities
up to the current turn of the conversation are identified.
% detected.
Next, paths 
%(facts) 
from our KG model involving these entities are extracted, and 
{\em RL agents} walk along these paths to candidate answers. The paths to walk on (actions by the agent) are decided according to predictions from a 
%randomly initialized 
{\em policy network}, which takes as input the conversational context and the KG paths. Aggregating {\em end points} of walks by the different agents leads to the final answer.
Upon observing this answer, the user issues a \textit{follow-up question}.
A \textit{reformulation predictor} takes this <original question, follow-up question> sequence as input and outputs a reformulation likelihood.
Parameters of the policy network are then updated in an online manner using rewards that are based on this likelihood.
Context entities are reset at the end of the conversation, but the policy parameters continue to be updated as more and more conversations take place between the user and the system.
Sec.~\ref{sec:startpt} through \ref{sec:ansgen} 
describe \conquer in detail.

	% !TEX root = ../2021-sigir-fp-conquer.tex
\section{Detecting context entities}
\label{sec:startpt}

% We are now in a position to describe our technique,
% that consists of two major blocks:
% start point selection, and performing the walk.
% , and aggregating answers.

% %Note that to faithfully
% %	address the problem of ConvKGQA, we cannot 
% % We do not assume entities to be given, nor do we rely on entity-linking tools~\cite{ferragina2010tagme,hoffart2011robust,van2020rel}) .
% %like
% %	TagME~\cite{ferragina2010tagme} or AIDA~\cite{hoffart2011robust}
% %	to get the salient entities (such tools need a well-formed sentence; 
% Entity-linking would be
% % notoriously
% extremely poor on % the kinds of
% incomplete utterances;
% % for example, there is
% (no hope for disambiguating ``Marvel'' in $Q_2$ correctly to \textit{Marvel Cinematic Universe} in contrast to \textit{Marvel Comics} or \textit{Marvel Studios}). 
% %
% Since the total length of this queue can get quite large over the duration of the conversation, the top-$k$ nodes are fetched. 

\begin{comment}
Selecting
% start points
context entities $E^{cxt}$ is a substitute for entity detection
in conversational questions, as off-the-shelf NED tools
cannot work without sufficient context. The goal is to
find good entity candidates (possibly null) in a partial utterance like
\utterance{What was the next from Marvel?}
that would potentially be points from where the RL agent would
start its walk at this turn.
\end{comment}
Throughout a conversation $C$, we maintain a set of
context entities $E^{cxt}$ that reflect the user's topical focus and intent. For full-fledged questions this would call for
Named Entity Disambiguation (NED), linking entity mentions onto KG nodes~\cite{shen2014entity}. There are many
methods and tools for this purpose, 
%(e.g., TagMe, AIDA, ... 
%%%GW: lists more (incl. NLP/IR big-shots) or none
including a few that are geared for very short inputs like
questions~\cite{sawant2013learning,piccinno2014tagme,li2020efficient}.
However, none of these can handle contextually incomplete and colloquial utterances that are typical for follow-up questions in
a conversation, for example: \utterance{What came next from Marvel?} or
\utterance{Who played his girlfriend?}.
The set $E^{cxt}$
% Start points
is comprised of
% identified as
the relevant KG nodes
% from the KG-context
for the conversation so far
and is created and maintained as follows.
% where the KG-context is defined as follows.

% \conquer maintains
% a set of KG nodes that have been relevant to the conversation
% so far, 
The set $E^{cxt}$
emanates from the question keywords
%and 
% This set
%is null at the start of the conversation.
%It 
and
is initialized by running an NED tool on the \textit{first question}
$q_1$, which is almost always well-formed and complete.
% rishi: not sure if we need the line below, could raise some questions
% As a special case, $E^{cxt}$ could be empty, but this should be rare.
Further turns in the conversation incrementally
augment the set $E^{cxt}$.
Correct answers for questions could qualify for $E^{cxt}$
and would be strong cues for keeping context.
%because follow-up questions may ask about them without explicit referencing.
However, they are not considered by \conquer: 
the online setting that we tackle does not have any knowledge of ground-truth answers and would thus have to 
indiscriminately pick up both correct and incorrect answers.
Therefore, \conquer considers only entities derived from
user utterances. %, and related ones from KG neighborhoods.

Let $E^{cxt}_{t-1}$ denote the set of context entities up to turn $t-1$.
% Neighbors of nodes (usually the $1$-hop neighborhood suffices as conversations operate in localized KG contexts) in 
Nodes in the neighborhood $nbd(\cdot)$ of $E^{cxt}_{t-1}$ form
the \textit{candidate context} for turn $t$ and are subsequently scored for entry into $E^{cxt}_t$. 
In our experiments, we restrict this to 1-hop neighbors,
which is usually sufficient to capture all relevant cues.
% perhaps the text below sounds defensive
% The reason why we consider that
% a $2$-hop neighborhood is less practical here is two-fold:
% (i) the agents will walk further from the start points
% selected eventually, so can anyway reach the $2$-hop neighborhood
% of $X_{t-1}$;
% (ii) a conversation usually consists of a sequence of
% coherent and sequential information needs, and a direct multi-hop intent
% ($3$- or more hops)
% in a follow-up question would most likely be broken down into
% conceptually simpler ones;
% (iii) scoring all nodes in the $2$-hop neighborhood for all points in $X_t$
% could lead to an explosion in the candidates by about
% three to four orders of magnitude~\cite{christmann2019look}.
For scoring candidate entities $n$ for question $q_{t}$ at turn $t$,
\conquer computes four measures for each 
%candidate node 
$n \in nbd(e | e \in E^{cxt}_{t-1})$:
\squishlist
	\item \textbf{Neighbor overlap:} This is the number of nodes in $E^{cxt}_{t-1}$ from where $n$ is reachable in one hop,
%	($n$ is in immediate vicinity of the previous context):
	the higher the better. Since this indicates high connectivity to $E^{cxt}_{t-1}$, such nodes are potentially good candidates. The number is normalized by the cardinality $|E^{cxt}_{t-1}|$, to produce $overlap(n) \in [0, 1]$.
	%MK: lexical match is done with Jaccard overlap
 	\item \textbf{Lexical match:} This is
 	% measured by
 	the Jaccard overlap between the set of words in the node label of $n$ and all words in $q_{t}$ (with stopwords excluded): $match(n) \in [0, 1]$.
%  	indicating whether the node label of $n$ contains any of the keywords in $Q_i$ (\phrase{marvel} in $Q_2$
% 	matches \struct{Marvel Cinematic Universe}).
	% Since we are looking for named entity candidates, a sophisticated semantic matching with word embeddings could lead to lots of out-of-vocabulary cases; surface forms often suffice for entity names. Note that searching the existing context for keyword matches is much more guided than an NED tool that tries to disambiguate over the whole of the KG $K$.
	\item \textbf{NED score:} 
	%We believe that good NED systems can still be useful where we have some context, and use the confidence scores from off-the-shelf systems as one of our signals. 
	Although full-fledged NED would not work well for incomplete and colloquial questions, NED methods can still give useful signals. We run an off-the-shelf tool, which we provide with richer context by concatenating the current with the 
	previous questions as input.
%	and consider their
%	confidence scores in the returned entities, normalized to $[0,1]$.
	We consider its normalized confidence score, $ned(n) \in [0, 1]$, but only if the returned entity is in the candidate set $nbd(e | e \in E^{cxt}_{t-1})$; otherwise $ned(n)$ is zero. This can be thought of as NED restricted to
	the neighborhood of
	the current context $E^{cxt}$ as an entity repository.
	%, instead of the whole KG.
	%Confidence scores from NED are usually already between zero and one, such that $ned(n) \in [0, 1]$. 
%	To provide the NED tool with richer context, we concatenate the current with the 
	%previous questions. % $Q_{i-1}$ and $Q_i$.
%	previous questions as its input.
	% \GW{only preceding, not all previous ones - right?}
	% rishi: yes, only preceding worked best among some other variants
	%mk: now for elq it is all previous ones
    \item \textbf{KG prior:} Salient nodes in the KG, as measured by the number of facts they are present in as subject, are indicative of their importance in downstream tasks like QA~\cite{christmann2019look}. A prior on this KG frequency often helps discriminate obscure nodes from prominent ones.
    % (\struct{Marvel Cinematics Universe} with $72$ triples would be considered more likely here $\struct{Marvel Music}$ with $8$ triples).
    We clip raw frequencies at a factor $f_{max}$, and normalize them by $f_{max}$  
    %by the number of all triples in $K$, 
    to yield $prior(n) \in [0, 1]$.
\squishend

These four scores are linearly combined with hyperparameters $h_1, \ldots, h_4$, such that $\sum_{i = 1}^{4} h_i = 1$, to compute the {\em context score}: % $cxt(n) = h_1 \cdot overlap(n) + h_2 \cdot match(n) + h_3 \cdot ned(n) + h_4 \cdot prior(n)$.
\vspace*{-0.3cm}
\begin{equation} \small
\vspace*{-0.3cm}
 	cxt(n) = h_1 \cdot overlap(n) + h_2 \cdot match(n) + h_3 \cdot ned(n) + h_4 \cdot prior(n)
\vspace*{-0.3cm}
\end{equation}
If score $cxt(n)$ is above a specified threshold $h_{cxt}$, 
then $n$ is inserted 
into the set of context entities $E^{cxt}_t$.
%list
%%%GW: why list? it was referred to as a set so far?
%% priority queue
%of context entities $E^{cxt}_t$ for conversation $C$ so far.
%
% with priority value $start(n)$.
% The priority values of nodes existing in the queue, $\{n_{old}\}$, are adjusted by decaying their previous values by multiplying them with  weights $t_{entry}(n_{old})/t$, where $t_{entry}(n_{old})$ is the turn where $\{n_{old}\}$ was first inserted into the queue.
% Due to the nature of conversations, the entities originating from the first turn (signifying the topic of the conversation) and the previous turn (immediate context) are always more important than others~\cite{qu2019bert,christmann2019look}. As a result, we multiply them with one so that they stay high priority in the queue.
Hyperparameters $h_1, \ldots, h_4$ and $h_{cxt}$ are tuned on a development set. % Finally, the selected
Entities in $E^{cxt}_t$
% from the priority queue
are passed on
% to the next phase,
%, for this turn $t$,
as
start points 
% $\{sp_t\}_{i=1}^k$
for RL agents to walk from (Sec.~\ref{subsec:rltrain}).

 	\vspace*{-0.2cm}
\section{Learning from reformulations}
\label{sec:method}

% everything about rl goes here:
% rl basics, training, answering

\subsection{RL Model}
\label{subsec:rl-map}

%\begin{figure} [t]
%	\centering
%	\includegraphics[width=0.5\columnwidth]{images/rl.png}
%	% \vspace*{-0.7cm}
%	\caption{RL basics.}
%	\label{fig:rl}
%	\vspace*{-0.3cm}
%\end{figure}

% \GW{this first paragraph sounds like an amateur's intro to RL -- I suggest dropping it, not needed for a scientific paper submitted to experts}\\
% rishi: dropped
% %Basic ideas behind the mapping of our ConvQA problem with reformulations to an RL setup are presented below.
% %%The goal is to learn a similarity matching between questions and KG facts.
% %In Reinforcement Learning, an  \textit{agent} is interacting with an \textit{environment}, moving between \textit{states} by selecting an \textit{action} in each step and receiving \textit{rewards} from the environment for the action taken.
% %Based on sequences of these actions and rewards, the agent learns an optimal policy for selecting the optimal action in a particular state. 
% %In our setting the environment consists of the user's questions and the KG. The agent traverses the KG to 
% Key concepts for mapping ConvQA to an RL task are
% \textit{an agent, the environment, a state, an action, a policy}, and \textit{a reward}.
% An agent, situated in a given state, takes an action, after which it receives a reward from the environment
% along with a new state to which it moves.
% Based on sequences of these actions and rewards,
% the agent learns an optimal policy for selecting an action in a given state,
% for achieving some given goal.
% %by which it chooses further actions towards achieving its goal.
% % transition function?

The goal of an RL agent here is to learn to answer conversational questions correctly.
%while needing minimal reformulations.
The user (issuing the questions) and the KG jointly represent the environment.
The agent walks over the knowledge graph $K$,
where entities are represented as nodes and predicates as path labels (Sec.~\ref{subsec:kg-model}).
% , as  explained
% in the \conquer KG model (Sec.~\ref{subsec:kg-model}).
An agent can only start and end its walk at entity nodes (Fig.~\ref{fig:qualifier}),
after traversing a path label.
%Every turn corresponds to a one-hop walk over $K$ by each agent, where the action (i.e., the edge label)
%is using a learnt policy.
%Start points are chosen as discussed in the previous section; end points are candidate answers.
This traversal can be viewed as a Markov Decision Process (MDP), where individual parts $(\mathcal{S}, A, \delta, R)$ are defined
as follows (adapted from~\cite{das2018go}):
\begin{itemize}
	\item \textbf{States:} 
	A state $s \in \mathcal{S}$ is represented by $s = (q^{cxt}_t, q_t, e^{cxt}_t)$, where
	$q_t$ represents the question at turn $t$ (new intent or reformulation), 
	$q^{cxt}_t \in Q_{t}^{cxt}$ captures a subset of the previous utterances as the (optional) context questions and $e^{cxt}_t \in E_{t}^{cxt}$ is one of the
	% salient conversational
	context entities for turn $t$ that serves as the starting point for an agent's walk.
	%where $H_t$ represents the conversation history, 
	%$Q_t$ is the question at turn $t$ and $e^{cxt}_t \in E_{t}$ is one of the available start points  at turn $t$.
	\item \textbf{Actions:} The set of actions $A_{s}$ that can be taken in state $s$ consists of all outgoing paths of the entity node $e^{cxt}_t$
	in $K$, so that
	$A_{s} =\{p | \langle e^{cxt}_t, p, e^{ans} \rangle \in K\}$. 
	%	$A_{s} =\{e^{cxt}_t, p | \langle e^{cxt}_t, p, e^{ans} \rangle \in K\}$. 
	End points of these paths are candidate answers $e^{ans}$.
	\item \textbf{Transitions:} % The environment evolves deterministically.
	The transition function $\delta$ updates a state
	% the next entity node $e^{ans}$ along with the next question, and
	% (optional) context questions for the next turn:
	to the agent's destination entity node $e^{ans}$ along with the follow-up question and (optionally) its context questions; 
	% updated conversational utterances
	% and the next question:
	$\delta: \mathcal{S}\times A \rightarrow \mathcal{S}$ is defined by $\delta(s,a) = s_{next} = (q^{cxt}_{t+1}, q_{t+1}, e^{ans})$.
% elements of the state remain unchanged.
%	\item \textbf{Environment:} The environment is modeled as the set of questions $Q_i$ and the processed KG $\mathcal{K}$.
%	\item \textbf{Episode:} A learning episode spans a specific intent $I$ in the conversation. It ends when a question with a new information need is encountered (no new reformulations for $I$).
	\item \textbf{Rewards:} The reward depends on the next user utterance $q_{t+1}$. If it expresses a new intent, then reward $R(s, a, s_{next})=1$. If $q_{t+1}$ has the same intent, then this is a reformulation, making $R(s, a, s_{next})=-1$.
	%, otherwise $R(s, a, s_{next})=1$. 
	%Future rewards are discounted with a factor $\gamma$ so that the agent selects actions to maximize the total of the discounted rewards it receives over an episode. This ensures that the learning goal is to get the answer correct in as few turns as possible. We also experiment with %\textit{reward shaping}~\cite{qiu2020stepwise,lin2018multi}, where the reward is not binary but directly the likelihood of non-reformulation ($1 - $ probability of reformulation).
\end{itemize}

%MK: the following part could be shortened a bit
While we know deterministic transitions inside the KG through nodes and path labels, users' questions are not known upfront. So we use
a \textit{model-free algorithm} 
% MK: will search for a citation 
that does not require an explicit model of the environment~\cite{sutton2018reinforcement}. Specifically, we use Monte Carlo methods that rely on explicit trial-and-error \textit{experience}: we learn from sampled state-action-reward sequences from actual or simulated interaction with the environment.
% and learn from sample experience, 
%  %in form of episodes 
% like it is done in Monte Carlo methods.
% \GW{Monte Carlo methods is a very generic, high-level term, not informative here; can we give a reference for model-free RL?}
% rishi: please see rephrasing above
Since questions can be arbitrarily formulated on any of several topics, 
our state space is
% would be
%infinitely large.
unbounded in size.
% if arbitrary conversations were possible,
% since arbitrary 
Thus, 
it is not feasible to learn transition probabilities between states. Instead, we use a \textit{parameterized policy} that learns to capture similarities between the question (along with its conversational context) and the KG facts. 
% \GW{model-free vs. parameterized -- sounds like a contradiction? clarify!}
% rishi: model-free because we do not try to find an explicit environment model (difficult in our case -- the user is a dynamic factor)
% and parameterized policy because we cannot hope to build a state transition probability matrix - there is an unquantifiable number of states due to the (question, context, node) tuples
% so it is not really a contradiction
The parameterized policy is manifested in the weights of a neural network, called the \textit{policy network}. When a new question arrives, this policy can be applied by an agent to
% take a path that is most likely to lead to the
reach an answer entity: this is equivalent to the agent following a path predicted by the policy network.

\vspace*{-0.2cm}
\subsection{RL Training}
\label{subsec:rltrain}

 \begin{figure} [t]
	\centering
	\includegraphics[width=\columnwidth]{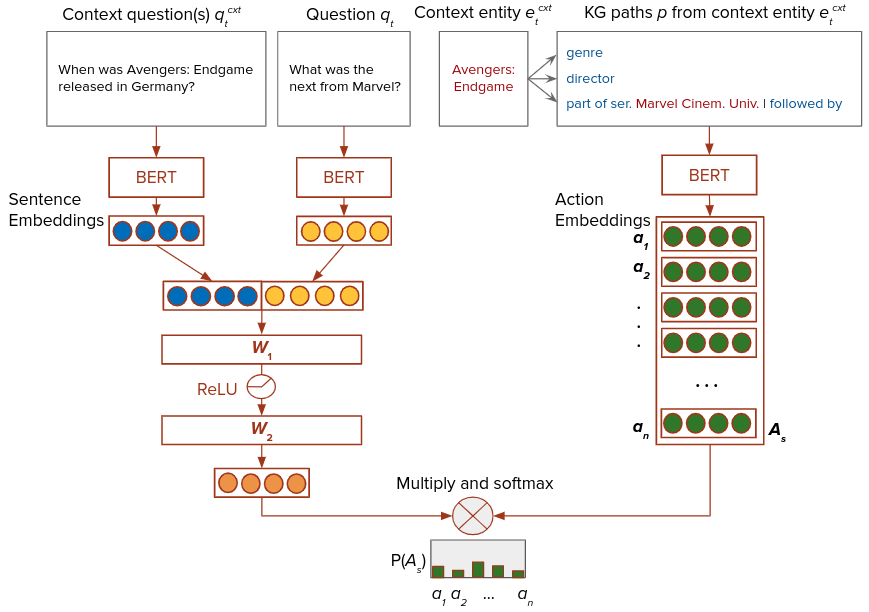}
	\vspace*{-0.7cm}
	\caption{Architecture of the policy network in \conquer.}
	\label{fig:policy-network}
	\vspace*{-0.3cm}
\end{figure}

%\vspace*{-0.3cm}
% \utterance{How to write a question?}
% \phrase{Spiderman 2}
% \struct{Avengers: Endgame}
% We use a policy network to learn our parameterized policy
% An alternative is to % improving the policy directly is to 
% learn a value function: this 
% function
% makes an estimate of the future rewards that the agent will obtain from a particular state onward, when following a certain policy. However, 
%MK: maybe leave out this first sentence here for space reasons?!
Using a policy network has been shown to be more appropriate for KGs due to the large action space~\cite{xiong2017deeppath}. The alternative of using value functions (e.g. Deep Q-Networks~\cite{mnih2015human}) may have poor convergence in such scenarios. 
% \GW{avoid such long footnotes in the middle of some argument; either it is secondary, then drop, or it is important, then integrate into main text!}
%rishi: now in main text, but briefly
To train our
% policy
network,
we apply the policy-gradient algorithm REINFORCE with baseline~\cite{williams1992simple}. As baseline, we use the average reward over several training samples
for variance reduction.
 % for training. 
The parameterized policy $\pi_{\boldsymbol{\theta}}$ takes information about a state as input and outputs a probability distribution over the available actions in this state. 
Formally: $\pi_{\boldsymbol{\theta}}(s) \mapsto P(A_{s})$.
%Following previous works, we use a two-layer feed forward network with a ReLU activation. %as policy network.

Fig.~\ref{fig:policy-network} depicts our policy network which contains a two-layer feed-forward network with non-linear ReLU activation. 
The policy parameters $\boldsymbol{\theta}$ consist of the weight matrices $\boldsymbol{W_1}$, $\boldsymbol{W_2}$ of the 
feed-forward layers.
%Sentence embeddings for 
%the current question and, optionally, the conversational context, are obtained from a BERT-base model by averaging over all hidden layers.
%The input consists of the the embeddings of the current question $\boldsymbol{Q_t}$ $\in \mathbb{R}^d$ which can optionally be concatenated with the embeddings of previous conversational context  $\boldsymbol{H_t}$ $\in \mathbb{R}^d$. 
Inputs to the network consist of
embeddings of the current question $\boldsymbol{q_t} \in \mathbb{R}^d$ (\utterance{What was the next from Marvel?}), optionally prepended with some context question embeddings % history
% the embeddings of the conversational context
$\boldsymbol{q^{cxt}_t} \in \mathbb{R}^d$ (like \utterance{When was Avengers: Endgame released in Germany?}).
% conversational context is very broad
% with embedding dimension $d$. %, concatenated with the embedding of the ).
% Thus, the input is of size $\mathbb{R}^{2d}$.
We apply % an uncased BERT-base
a pre-trained BERT model to obtain these embeddings by averaging over all hidden layers and over all input tokens.
Context entities $E^{cxt}_t$ (e.g., \struct{Avengers: Endgame}) are the starting points for an agent's walk
% are determined upfront
and are identified as explained in Sec.~\ref{sec:startpt}. 
% One possible starting point is ``Avengers Endgame'' in our example.
We then retrieve all outgoing paths for these entities from the KG.
%There can be several startpoints per question. For each possible startpoint, we query its one hop neighborhood.
%better in experiment details?
%and retrieve up to 1000 paths from our KG. 
% rishi: probably breaking the flow
% Note that we squeezed in qualifiers in our paths, as described in section~\ref{subsec:kg-model}. Thus, we are able reach entities which are likely two hops away in other KG representations.
%Thus, the set of actions available in state $s$ is of size $|A_{s}| = 1000$.
An action vector $\boldsymbol{a} \in \boldsymbol{A_{s}}$ consists of the embedding of the 
%context entity along with the respective path: $(e^{cxt}_t; p) \in \mathbb{R}^d$. 
 respective path $p$ starting in $e^{cxt}_t$, $\boldsymbol{a} \in \mathbb{R}^d$. 
These actions are also encoded using BERT.
%the same BERT model that was used to embed $q_t$ and $q^{cxt}_t$.
The final embedding matrix $\boldsymbol{A_{s}} \in \mathbb{R}^{|A_{s}| \times d}$ consists of the stacked action embeddings. %for the respective startpoint. 
The output of the policy network is the probability distribution $P(A_{s})$, that is defined as follows:
%$$\pi_\boldsymbol{\theta}(a_j | s) = \sigma(\boldsymbol{A_{s}}(\boldsymbol{W_2} \textrm{ReLU} (\boldsymbol{W_1}([H_t, Q_t]))))$$
\begin{equation}
	\vspace*{-0.1cm}
    P(A_{s}) = \sigma(\boldsymbol{A_{s}} \times (\boldsymbol{W_2} \times ReLU (\boldsymbol{W_1} \times [\boldsymbol{q^{cxt}_t}; \boldsymbol{q_t}])))
    \label{eq:relu}
     % \vspace*{-0.1cm}
\end{equation}
where $\sigma(\cdot)$ is the softmax operator.
Then, the final action which the agent will take in this step is sampled from this distribution:
\begin{equation}
    \vspace*{-0.1cm}
    a = A_{s}^{i}, \ i \sim Categorical(P(A_{s}))
    \label{eq:action}
    % \vspace*{-0.1cm}
\end{equation}
%$a = A_{s}[a_{num}]$, where $a_{num} \sim Categorical(P(A_{s}))$  
%MK: maybe remove J(\boldsymbol{\theta}) part??
To update the network's parameters $\boldsymbol{\theta}$, the expected reward $J(\boldsymbol{\theta})$ is maximized over
each state and the corresponding set of actions:
% rishi: rephrase above if necessary
% each question along its conversational context in the training set,
% determining our current state:
%all questions and their follow-up requests in the dataset:
%J(\boldsymbol{\theta}) = \mathbb{E}_{Q_t, Q_{t+1}}[\mathbb{E}_{a \sim \pi_\boldsymbol{\theta}}[R(s,A_{s},s_{next}) | (Q_t, Q_{t+1})]]$$
\begin{equation}
     \vspace*{-0.1cm}
    J(\boldsymbol{\theta}) = \mathbb{E}_{s \in S}\mathbb{E}_{a \sim \pi_{\boldsymbol{\theta}}}[R(s,a,s_{next})]
    \label{eq:jtheta}
     % \vspace*{-0.1cm}
\end{equation}

\begin{algorithm} [h!]
%\begin{algorithmic}[1]
	\small
	\DontPrintSemicolon
	\SetAlgoLined
	\SetKwInput{Input}{Input}\SetKwInput{Output}{Output}
	\SetKwComment{Comment}{$\triangleright$\ }{}
%	\Input{a differentiable policy parameterization $\pi(a | s, \boldsymbol{\theta})$}
		\Input{question $q_t$, KG $K$,  step size ($\alpha > 0$), number of rollouts ($rollouts$), size of update ($batchSize$), entropy weight ($\beta$)}
		\Output{updated policy parameters $\boldsymbol{\theta}$}
%	Algorithm parameters: step size ($\alpha > 0$), number of rollouts ($rollouts >0$), update frequency ($ufreq > 0$), current step ($step >= 0)$\;
	\Comment{On initial call: Initialize $\boldsymbol{\theta}$ randomly} 
	%and initialize current step $step \gets 0$\;
	$q^{cxt}_t \gets loadContext(q_t)$\;
	$E_{t}^{cxt} = detectContextEntities(q^{cxt}_t, q_t)$ \;%\emptyset$\;
	%\ForEach{$n \in nbd(e | e \in E_{t-1})$}
	%{$cxt(n) = h_1 \cdot overlap(n) + h_2 \cdot match(n) + h_3 \cdot ned(n) + h_4 \cdot prior(n)$
%	\If{$cxt(n) > h_{cxt}$}
%	{$E_{t}.add(n)$}}
	\ForEach{$e^{cxt}_t \in E_{t}^{cxt}$}
    	{$\mathcal{P} \gets getKGPaths(e^{cxt}_t,K)$ \;
    	$s \gets (q^{cxt}_t, q_t, e^{cxt}_t)$ \;
    	$\boldsymbol{q_t} \gets BERT(q_t)$, $\boldsymbol{q^{cxt}_t} \gets BERT(q^{cxt}_t)$ \;
    	$\boldsymbol{A_{s}} \gets stack(BERT(p))$, where $p \in \mathcal{P}$ \;
    	$P(A_{s}) = \sigma(\boldsymbol{A_{s}} \times (\boldsymbol{W_2} \times ReLU (\boldsymbol{W_1} \times [\boldsymbol{q^{cxt}_t}; \boldsymbol{q_t}])))$ \;
    	$count \gets 0$ \;
    	\While {$count < rollouts$}
        	{$i \sim Categorical(P(A_{s}))$\;
        	$a \gets A_{s}^{i}$, where $a := p_i$ and $(e^{cxt}_t,p_i,e^{ans}) \in K$ \;
        	%where $a := (e^{cxt}_t,p)$ and $(e^{cxt}_t,p,e^{ans}) \in K$ \;
        %	$answer \gets e^{ans}$ \;
        %	$s_{next} \gets (q_{t+1}, q^{cxt}_{t+1}, e^{ans})$ where $a = (e^{cxt}_t,p)$ and $(e^{cxt}_t,p,e^{ans}) \in K$ \;
        	$q_{t+1} \gets getUserFeedback(e^{ans})$ \; 
        	$q^{cxt}_{t+1} \gets loadContext(q_{t+1})$ \;
        	$s_{next} \gets (q^{cxt}_{t+1}, q_{t+1},  e^{ans})$\;
        \lIf{$isReformulation(q_t, q_{t+1})$}
        	    {$R \gets -1$}
        	\lElse{$R \gets 1$}
        	$experience.enqueue(s,a, s_{next},R)$ \;
        	$count \gets count + 1$}
       % $step \gets step + rollouts$
    }
	\uIf{$|experience| >= batchSize$}
	 {%\Comment{retrieve $batchSize$ entries from experience buffer} 
    $updateList \gets experience.dequeue(batchSize)$\;
    $batchUpdate \gets 0$, $RList \gets getRewards(updateList)$ \;
    $\bar{R} \gets mean(RList)$, $\sigma_R \gets std\_dev(RList)$ \;
    \ForEach {$(s,a,s_{next}, R) \in updateList$}
           	{$H_\pi(\cdot, s) \gets - \sum_{a \in A_{s}} \pi(a|s) \cdot log \pi(a|s)$ \;
           	$R^* \gets \frac{R-\bar{R}}{\sigma_R}$\;
           	$batchUpdate \gets batchUpdate + R^* \frac{\nabla \pi(a | s, \boldsymbol{\theta})}{\pi(a | s, \boldsymbol{\theta})} + \beta  H_\pi(\cdot, s)$}
           	%, with $H_\pi(\cdot, s) = - \sum_{a \in A_{s}} \pi(a|s) \cdot log \pi(a|s)$}
    	$\boldsymbol{\theta} \gets \boldsymbol{\theta} + \alpha \cdot  batchUpdate$ \;
    %	$step \gets 0$
    }
    %$step \gets step + 1$\;
    \Return{$\boldsymbol{\theta}$}

%	   \eindent
%	   \eindent
% \vspace*{-0.3cm}
\caption{Policy learning in \conquer}
% \vspace*{-0.3cm}
\label{alg:conquer}
% \vspace*{-0.3cm}
\setlength{\textfloatsep}{0pt} 
\end{algorithm}

%For the second expectation
%We use the policy gradient algorithm REINFORCE for optimization. 
For each question in our training set, we do multiple \textit{rollouts},
%to estimate the stochastic gradient:
meaning that the agent samples multiple actions for a given state to estimate the stochastic gradient (the inner expectation in the formula above).
%to get a good estimate of the expected return for this state.
%the second expectation in the formula above.
Updates to our policy parameters $\boldsymbol{\theta}$ are performed in batches.
Each \textit{experience} of the form ($s, a, s_{next}, R$) that the agent has encountered is stored. A batch of experiences is used for the update,
%after seeing multiple training samples, 
performed as follows:
% according to this formula: % (several episodes).
\begin{equation}
     \vspace*{-0.1cm}
    \nabla J(\boldsymbol{\theta}) =  \mathbb{E}_\pi [\alpha \cdot  (R^* \cdot \frac{\nabla \pi(a| s, \boldsymbol{\theta})}{\pi(a| s, \boldsymbol{\theta})} + \beta \cdot H_\pi(\cdot, s))]
    \label{eq:gradj}
         \vspace*{-0.1cm}
\end{equation}
where $\alpha$ is the step size, $R^*$ is the normalized return, $\beta$ is a weighting constant for $H_\pi(\cdot, s)$~\cite{buck2018ask,das2018go}, that is an entropy regularization term: % defined as:
\begin{equation}
     \vspace*{-0.1cm}
    H_\pi(\cdot, s) = - \sum_{a \in A_{s}} \pi(a|s) \cdot log \; \pi(a|s)
    \label{eq:ent-reg}
    \vspace*{-0.1cm}
\end{equation}
which is added to the update to ensure better exploration and prevent the agent from getting stuck in local optima.

%MK: the two following lines could be dropped in case we need space:
Parameters $\boldsymbol{\theta}$ are updated in the direction that increases the probability of taking action $a$ again when seeing $s$ next time. 
The update is inversely proportional to the action probability to not favor frequent actions
% that do not necessarily give the highest rewards
(see~\cite{sutton2018reinforcement}, Chapter 13, for more details).
Finally, we normalize each reward by subtracting the mean and by dividing by the standard deviation of all rewards in the current batch update: $R^* = \frac{R-\bar{R}}{\sigma_R}$ to reduce variance in the update. 
%using the mean and standard deviation of all rewards in the current batch update: $R^* = %\frac{R-\bar{R}}{\sigma_R}$ to reduce variance in the update. 
%The increment is proportional to the reward $R$ multiplied with a vector, the gradient of the probability of taking the action actually taken divided by the probability of taking that action.  
%The vector is the direction in parameter space that most increases the probability of repeating the action $a$ on future visits to state $s$. The update increases the parameter vector in this direction proportional to the return and thus it favors actions that give high return.
%Furthermore, it is inversely proportional to the action probability to not give advantage to frequent actions which might not have highest return.
%TODO : We use normalized returns -> explain?
% Additionally, we add 
%MK: maybe put earlier to make references to it?
\textbf{Algorithm~\ref{alg:conquer}} shows
high-level pseudo-code for the policy learning in \conquer.
% In the following,
We now describe how we obtain the rewards used in the update.

% \setlength{\parskip}{0ex}
%\subsection{Applying the policy network}
%maybe only explain in experiment section?? anyway described similarly in workflow
%After a training phase, the policy network can be used to make predictions.
%Once the policy network has been trained with a large number of question-reformulation pairs, 
%During test time, we first extract the context entities as described in section~\ref{sec:startpt}. 
%Most times we have several of these and thus several starting points for the RL walk. Thus, multiple agent make steps on the KG in parallel based on the predictions of the trained policy network. Each agent takes the top $k$ predicted actions for the respective startpoint. 
%These are sorted in the answer aggregation step according to their individual scores and by the number of agents reaching them.

\subsection{Predicting reformulations}
\label{subsec:refpred}

%Describe everything about the reformulation prediction module here.
Each answer entity $e^{ans}$ reached by an agent after taking the sampled action is presented to the user. 
% The user can now react based on the shown answer and either issue a reformulation with the same intent or ask a follow-up question with a new intent.
An ideal user, according to our assumption, would ask a follow-up question that is either a reformulation of the same intent (if she thinks the answer is wrong), or an expression of a new intent (if the answer seems correct). 
%During training, we simulate the user.
%During training, 
This sequence of the original question and the follow-up is then passed
% along with the previous question
on to a reformulation detector to 
%The reformulation predictor
% that
decide whether the two questions express the same intent. % or not. 
We devise such a predictor by fine-tuning
% an uncased 
% \GW{what is "uncased"? do you mean that upper-case/lower-case is disregarded?}
% rishi: yes upper/lower case is disregarded, but may be this is better for the experiments... but this specificity allows us to use "the same BERT model" later
% bert-base-uncased: https://huggingface.co/bert-base-uncased
% bert-base as opposed to bert-large
% BERT-base model\footnote{\url{https://huggingface.co/bert-base-uncased}}
a BERT model
for sentence pair classification~\cite{devlin2018bert} % for this task
on a large set of such question pairs.
% \footnote{If real reformulations are unavailable, one could explore alternatives like thresholding on sentence-sentence similarities using off-the-shelf sentence embedding models, using pre-trained sentence-sentence models for entailment prediction, or fine-tuning similar models with paraphrasing resources.}
% \GW{again, avoid lengthy footnotes! either drop or integrate into main text! entailment prediction needs reference, for IR folks}
% rishi: dropped
Based on this prediction (reformulation or not), we deduce if the generated answer entity $e^{ans}$ was correct (no reformulation) or not (reformulation). 
The agent then receives a positive reward (+1) when the answer was correct and a negative one (-1) otherwise. % (reformulation).

\section{Generating answers}
\label{sec:ansgen}

The
% trained
learned
% \GW{better say "learned" rather than "trained" to hint at the RL flavor?}
%\conquer model 
policy 
can now be used to generate answers for conversational questions. %using its RL capability. 
This happens in two steps: i) selecting actions by individual agents to reach candidate answers and ii) ranking the candidates to produce the final answer.
%Note that the training can continue in an online fashion.

\myparagraph{Selecting actions}
% In the training phase, we made predictions to update our policy parameters.
% Now we can apply the trained policy network to answer user requests. % without further updating the network.
% As before, the first step consists of  extracting the context entities.
Given a question $q_i$ at turn $t$, the first step is to extract the context entities $\{e^{cxt}_t\}$.
%as described in section~\ref{sec:startpt}. 
%During testing, we first extract the context entities as first described in section~\ref{sec:startpt}. 
% Most times we have
Usually there are several of these and, therefore, several starting points for RL walks.
Multiple agents 
%take steps on 
traverse
the KG in parallel, based on the predictions coming from the trained policy network. 
% During training the agents sample actions from the probability distribution (see line x in algorithm~\ref{alg:conquer}) to ensure exploration. At answering time, each agent greedily takes the top-$1$ predicted actions.
Each agent takes the top-$k$ predicted actions ($k$ is typically small, five in our experiments) from the policy network greedily (no explorations at answering time).
%MK: I think the following is not really necessary:
%Note that action prediction varies depending upon the context entity $e^{cxt}_t$, as each has different outgoing paths $\{p\}$, and both the entity and the paths are inputs to the policy network.
% rishi: top-k is tricky here, an agent can only take one action by definition
%MK: lets talk about top-k again, this is actually what is done
% step we need to decide which answer to present to the user
% we can still get top-5 by aggregation (ranking)
% for the respective startpoint to find the correct answers.

\myparagraph{Ranking answers}
%During training the entity reached when following the sampled action is shown to the user so that she can provide the system with feedback. 
%During answering time, we are displaying a short list of most probable answers to the user
%Since we have multiple agents making predictions and each agent retrieves the top-k actions we have multiple possible answers that could be returned to user. 
% While during training we require feedback for all sampled actions, at answering time \conquer should display the best possible answer to the user only.
% Thus, the answer entities selected by the agents need to be sorted.
Agents land at end points after following actions predicted for them by the policy network. These are all candidate answers $\{e^{ans}\}$ that need to 
% End points of walks by the multiple agents are all candidate answers, and need to
be
% aggregated or
ranked. % before presenting the best one to the user.
We interpret the
% individual
probability scores coming from the network (associated with the predicted action) as a measure of the system's confidence in the answer and use it as our main ranking
% is the main sorting
criterion. 
When multiple agents land at the same end point, we use this to boost scores of the respective candidates by adding
scores for the individual actions.
Candidate answers are then ranked by this final score, and the top-$1$ entity is shown to the user. % ($k$ is typically small). 
% However, we want to boost the score of answers that have been reached by several agents. In our example, \struct{Spider-man} can be reached via a walk from \struct{Marvel Cinematic Universe} and from \struct{Avengers Endgame}. This could indicate that it is more likely to be correct. We therefore add the scores coming from these two predictions.
%in the answer aggregation step according to their individual probability scores coming from the policy network and by the number of agents reaching them.

%The workflow is very similar as before, except that we now do not update the policy network anymore and

%As before, the first step consists of  extracting the context entities as described in section~\ref{sec:startpt}. 
%During testing, we first extract the context entities as first described in section~\ref{sec:startpt}. 
%Most times we have several of these and thus several starting points for the RL walk. Thus, multiple agent make steps on the KG in parallel based on the predictions coming from our trained policy network. 
%While during training the agent samples an action to ensure exploration, each agent greedily takes now the top $k$ predicted actions for the respective startpoint. 
%The entities selected this way are sorted in the answer aggregation step according to their individual probability scores coming from the policy network and by the number of agents reaching them.
	% !TEX root = ../2021-sigir-fp-conquer.tex
\section{Benchmark with Reformulations}
\label{sec:benchmark}

None of the popular QA benchmarks,
like WebQuestions~\cite{berant2013semantic}, 
ComplexWebQuestions~\cite{talmor2018web},
%SQuAD, 
QALD~\cite{usbeck20189th}, LC-QuAD~\cite{dubey2019lc,trivedi2017lc}, or CSQA~\cite{saha2018complex}, contain sessions with questions with reformulations by real users.
The only publicly available benchmark for ConvQA over KGs
based on a real user study
%thus adapted a very recent dataset for ConvQA over KGs 
is 
\convquestions~\cite{christmann2019look}.
% (the other ConvQA benchmark for KGs, CSQA~\cite{saha2018complex}, is created semi-synthetically using templates) 
% where questions span five domains: Movies, TV Series, Books, Music, and Soccer
% ({\small\url{https://convex.mpi-inf.mpg.de/}}).
% all other URLs are in normal sized font
% ({\url{https://convex.mpi-inf.mpg.de/}}).
%with an interactive user study to build our evaluation resource (\convquestionsref) as follows. 
%However, this is still not suited for our setting, as there
%are no system-generated answers and subsequent reformulations for user utterances.
% , which are needed in our setting to trigger 
% genuine reformulations.
% \GW{check this wording! the point should be made clear here, but in a single sentence!}
% \GW{or could simply say that convquestions lacks reformulations for incorrect answers ???}
Therefore, we used conversation sessions in \convquestions as input to % conducting
our own user study % as follows.
to create our benchmark % resource
\convquestionsref.
\myparagraph{Workflow for user study} 
%The goal was to collect true reformulations for each existing information need in \convquestions and not a static set of paraphrases.
Study participants interacted with a baseline ConvQA system. 
In this way, we were able to collect real reformulations issued in response to seeing a wrong answer, rather
than static paraphrases.
%This was achieved by letting the study participants interact with a baseline ConvQA system.
To create such a baseline system we trained our randomly initialized policy network with simulated reformulation chains.
\convquestions
% comes with pre-compiled interrogative paraphrases for questions
comes with one interrogative paraphrase per question,
% templates
% no templates here,
which could be viewed as a weak proxy
% source of
for reformulations. 
A paraphrase was triggered as reformulation in case of a wrong answer during training.
% The train set of $6,720$ conversations in \convquestions was used in the study. 
%For our user study, we hired
%$30$ students (Computer Science undergraduates). 
%Users interacted with the above baseline ConvQA model. % (as outlined above).

The conversations shown to the users were topically grounded in
the $350$ seed conversations
% those
in \convquestions.
% : 
%there are 
% $350$ {seed} conversations 
%(available at \url{https://convex.mpi-inf.mpg.de/}) that 
% were used to generate % the original dataset.
% \convquestions.
%each question had one paraphrase
% \GW{what is "the original dataset"? the initial questions, one per session ??? clarify and change wording !!!}
Since we have paraphrases for each question, we also use the conversations where the original questions are replaced by their paraphrased version. This way we obtain
%As each question was paraphrased once, we unfolded this into
$700$ conversations, each with $5$ turns.
%We separated out the paraphrases to create $350 \times 2 = 700$ conversations. Recall that each conversation has five questions, i.e., five intents.
%
%For each of these $700$ conversations, we show the original question from the benchmark to the user, along with the answer generated on-the-fly by the baseline model. 
Users were shown follow-up questions in a given conversation interactively, one after the other, along with the answer coming from the baseline QA system.
% a question and the answer from the baseline QA system.
%Whenever this is wrong (again, verifiable through the availability of gold answers in \convquestions), 
For wrong answers, 
the user was prompted to reformulate the question 
up to four times if needed.
In this way, users were able to pose
%realistic 
reformulations based on previous wrong answers and the conversation history. 
%The reformulation is again passed on to the backend system, and an answer returned to the user. If the system response is correct, the prompt moves on to the next intent in the conversation sequence. If not, the user is asked to reformulate again -- this process repeats up to a maximum of four times (simulating abandonment of this intent in frustration), after which the prompt automatically takes her to the next intent. 
%By having humans in the loop, these questions were true reformulations in response to seeing the wrong answer, rather than static
% templates for paraphrasing as in the original \convquestions sessions.
%paraphrases.
%anymore, as the users observe a system-generated answer before formulating the follow-up question. 
Note that the baseline QA system often gave wrong answers,
as our RL model had not yet undergone full training.
This provided us with a challenging stress-test benchmark,
where users had to issue many reformulations.
%Also note that at this stage, the system makes a lot of mistakes as the RL model has still had very little training -- this is actually a blessing in disguise as this allows us to collect a large number of realistic reformulations and create a rich resource. In rare cases when the system response is correct,  At all stages, the user is shown the entire conversational context including all previous intents, reformulations and system answers. 
% \GW{check if all this wording keeps the proper meaning, but should be much shorter!}

\myparagraph{Participants} % and task design} 
We hired
$30$ students (Computer Science graduates) for the study. 
Each participant annotated about $23$ conversations.
%($23 \times 30 = 690$, the remaining ten were randomly assigned to the users). 
The total effort required 
%$210$ hours (about nine hours per user). 
%9 hours per user (plus a 2-hour briefing), 
7 hours per user (including a 1-hour briefing),
and each user was paid 10 Euros per hour
(comparable to AMT Master workers).
%Participants were paid at the rate of 10 Euros an hour, that is slightly higher or comparable to an average Master worker at Amazon Mechanical Turk. 
%Respecting pandemic regulations, all annotations were performed over individual Zoom sessions. 
%For quality control, all browser activity was monitored by the authors; recordings were deleted upon completion of the experiments to respect the privacy of the annotators. 
%Annotators were briefed about the task guidelines in a two-hour introductory session, where they were explained necessary details about KGs, QA, ConvQA and reformulations. 
%All browser activities were recorded, with explicit user consent,
%and 
The final session data was sanitized to comply with 
privacy regulations. 
% \GW{not sure if we need to say all this}
\begin{comment}
We wanted to make these reformulations as natural as possible (an analysis of the collected data is provided in Table~\ref{tab:qualitative}), and hence abstained from detailed instructions. Rather, we showed users samples of conversations with reformulations, one from each domain. Users could already understand the nature of expected questions when they were shown the original question from the \convquestions benchmark, and hence did not formulate irrelevant or subjective questions that would be out of scope for the underlying KG. To allow for maximum creativity (equivalent to syntactic diversity), each user was assigned a mix of conversations from the five domains. It may be possible that if a user annotates all movie conversations, she runs out of new formulations: this would be rather detrimental for a learning agent.
\end{comment}
%%%GW: way too much detail, puts a lot of emphasis on this study -- however, this is not where the paper's main contributions lie
%Table~\ref{tab:qualitative} gives statistics about the collected types of reformulations, and Table~\ref{tab:reformulations} shows
%examples.

\myparagraph{Final benchmark} 
The final \convquestionsref benchmark was compiled as follows.
Each information need in the $11.2k$ conversations from \convquestions
%consists of the $11.2k$ conversations coming from the previous \convquestions collection 
%where each information need 
is augmented by the reformulations we collected. % respectively.
This resulted in a total of $262k$ turns
with about $205k$ reformulations.
%The reformulations from our user study 
%were compiled into the \convquestionsref benchmark as follows. %\convquestions generated its original $11.2k$ conversations by creating question sequences from either a seed question or its paraphrase. Thus, each utterance in the original benchmark can be found among our $700$ annotated conversations. So we replace, or rather augment, each singleton intent (i.e., one utterance) in the original benchmark by the reformulations we collected. 
%For 
%Each of the $11.2k$ conversations in the previous
%\convquestions collection, we replace the
%paraphrase-based questions by the true reformulations from
%our user study, leading to a total of $262k$ turns
%with about $205k$ reformulations.
%This resulted in the final resource, \convquestionsref: still with $11.2k$ conversations (no entirely new conversations were created) but with $262k$ turns (\convquestions had only $56k$), out of which about $206k$ are reformulations. 
%Noting that a QA model that was \textit{always} wrong would lead to $56k \times 5 = 280k$ turns (note that users reformulated each intent up to a maximum of four times), we can roughly say that the baseline generated a correct answer about $6.4\%$ of the time. 
We followed \convquestions ratios for the train-dev-test split, leading to 
%was respected and is propagated to \convquestionsref: 
$6.7k$ training conversations and $2.2k$ each for dev and test sets. 
%Notably, seed conversations in \convquestions (carrying over to \convquestionsref) also maintained these train-dev-test splits: in other words, there was no leakage of any form from the train set to the dev or test sets.
%% Wherever applicable,
%We looked at a random sample of collected reformulations and categorized them into 6 different types
While participants could freely reformulate questions, we noticed different patterns in a random sample of 100 instances (see Table~\ref{tab:qualitative}).
Examples of reformulations are shown in Table~\ref{tab:reformulations}.
%Table~\ref{tab:qualitative} gives statistics about the collected types of reformulations, and Table~\ref{tab:reformulations} shows examples.
Reformulations had an average length of about $7.6$ words, %(relatively stable across turns), 
compared to about $6.7$ for 
the initial questions per session.
%
\begin{comment}
RL agents in \conquer always learned from the train sets only. This is necessary for our subsequent comparison with \convex, a SoTA ConvQA system over KGs. 
%%%GW: why would we need to say this here ???
Reformulations had an average length of about $7.6$ words (relatively stable across turns), compared to about \textcolor{red}{$6.4$ for original questions}.
% rishi: double check
This is because reformulations often involve adding disambiguating words and entities to clarify the context better. Representative examples of reformulations in \convquestionsref for each domain are shown in Table~\ref{tab:reformulations}. 
\end{comment}
%
The complete benchmark 
%, along with necessary markup (original question, reformulation, wrong answer, correct answer) 
is available at \url{https://conquer.mpi-inf.mpg.de}.
% Dictionaries on paraphrases for short text units that denote query predicates or target classes (e.g., ``cast of'', ``starring in'') were used in early work on QA over KGs
% \cite{fader2013paraphrase,dong2017learning,berant2014semantic}.
% %\textbf{Paraphrases.} Early work in KGQA leveraged paraphrases~\cite{fader2013paraphrase,dong2017learning,berant2014semantic} as additional learning signals, 
% %that are connected to reformulations in the sense that both refer to the same information need expressed in different ways. The subtle 
% %difference is that paraphrases are formulated blind; there is no system-in-the-loop
% %and the system-generated answer plays no role. However, the conversational setup with ad hoc, incomplete utterances coupled with the unique goal of 
% %reducing the number of turns taken to reach the correct answer, makes
% %traditional learning from paraphrases inapplicable to ConvKGQA.
% A key difference to our setting, though, is that paraphrase dictionaries are static and compiled from corpus frequency statistics, whereas reformulations in ConvQA
% % to deal with 
% involves incomplete context and ungrammatical phrases in a situative manner.
% % Paraphrases collected in 
% % a conversational setup~\cite{christmann2019look} can be useful for initialization, though.

\begin{table} [t] \small
	\centering
 	% \resizebox{\columnwidth}{!}{
		% \setlength{\tabcolsep}{0.5em} % for horizontal padding
		\begin{tabular}{l c}
			\toprule
			\textbf{Nature of reformulation}            &	\textbf{Percentage}	\\ \toprule
			Words were replaced by synonyms	            & 	$15\%$		        \\
			Expected answer types were added	        &   $14\%$			    \\ 
			Coreferences were replaced by topic entity	&   $24\%$		        \\ 			
			Whole question was rephrased	            & 	$71\%$	            \\ 
			Words were reordered	                    & 	$5\%$	           \\ 
			Completed a partially implicit question	    & 	$20\%$	            \\ \bottomrule
	\end{tabular}%}
%	\caption{Qualitative types of reformulations in \convquestionsref, from an analysis over a random sample of 100 cases. Numbers do not add up to 100 as each reformulation may belong to multiple types.}
		\caption{Types of reformulations in \convquestionsref. Each reformulation may belong to multiple categories.}
	\label{tab:qualitative}
	\vspace*{-0.7cm}
\end{table}

\begin{table} [t] \small
	\centering
    % \resizebox{\columnwidth}{!}{
		% \setlength{\tabcolsep}{0.5em} % for horizontal padding
		\begin{tabular}{l}
			\toprule
		    \textbf{Original question:} \utterance{in what location is the movie set?} \textbf{[Movies]}    \\
            \textbf{Wrong answer:} \struct{Doctor Sleep}                                                      \\
            \textbf{Reformulation:} \utterance{where does the story of the movie take place?}               \\ \midrule
    	    \textbf{Original question:} \utterance{which actor played hawkeye?} \textbf{[TV Series]}        \\
            \textbf{Wrong answer:} \struct{M*A*S*H Mania}                                                   \\
            \textbf{Reformulation:} \utterance{name of the actor who starred as hawkeye?}                   \\ \midrule
    	    \textbf{Original question:} \utterance{release date album?} \textbf{[Music]}                    \\
            \textbf{Wrong answer:} \struct{01 January 2012}                                               \\
            \textbf{Reformulation:} \utterance{on which day was the album released?}                        \\ \midrule    %       
    	    \textbf{Original question:} \utterance{what's the first one?} \textbf{[Books]}                  \\
            \textbf{Wrong answer:} \struct{Agatha Christie}                                                 \\
            \textbf{Reformulation:} \utterance{what is the first miss marple book?}                         \\ \midrule
            \textbf{Original question:} \utterance{who won in 2014?} \textbf{[Soccer]}                      \\
            \textbf{Wrong answer:} \struct{NULL}                                                            \\
            \textbf{Reformulation:} \utterance{which country won in 2014?}                                  \\ \bottomrule
	\end{tabular} % }
	\caption{Sample reformulations from \convquestionsref.}
	\label{tab:reformulations}
	\vspace*{-1cm}
\end{table}
	% !TEX root = ../2021-sigir-fp-conquer.tex
\section{Experimental framework}
\label{sec:exp-setup}

\subsection{Setup}
\label{subsec:setup}

\myparagraph{KG} We used the Wikidata NTriples dump from 26 April 2020 with about $12B$ triples. Triples containing URLs, external IDs, language tags, redundant labels and descriptions were removed, leaving us with about $2B$ triples (see \url{https://github.com/PhilippChr/wikidata-core-for-QA}). The data was processed according to Sec.~\ref{subsec:kg-model} and loaded into a Neo4j graph database (\url{https://neo4j.com}). 
All data resided in main memory (consuming about $35$ GB, including indexes) and was accessed with the Cypher query language. 
% (similar to SPARQL).
%for lookups. 

\myparagraph{Context entities} We used \elq~\cite{li2020efficient} to obtain NED scores.
% for detecting $E^{cxt}$.
%TagME~\cite{ferragina2010tagme} for NED scores for detecting $E^{cxt}$ (see Sec.~\ref{sec:startpt}).
%TagME's internal threshold was set to $0.3$, as tuned on our development set of $100$ manually annotated utterances.
% from \convquestions. 
Frequency clip $f_{max}$ was set to $100$. Parameters $h_1, h_2, h_3, h_4, h_{cxt}$ were tuned on our development set of $100$ manually annotated utterances and set to $0.1, 0.1, 0.7, 0.1, 0.25$, respectively (see Sec.~\ref{sec:startpt}).

\myparagraph{RL and neural modules} The code for the RL modules was developed using the TensorFlow Agents library (\url{https://www.tensorflow.org/agents}). When the number of KG paths for a context entity exceeded $1000$, a thousand paths were randomly sampled owing to memory constraints. All models were trained for $10$ epochs, using a batch size of $1000$ and $20$ rollouts per training sample. All reported experimental figures are averaged over five runs, resulting from differently seeded random initializations of model parameters. We used the Adam optimizer~\cite{kingma2015adam} with an initial learning rate of $0.001$.
The weight $\beta$ for the entropy regularization term is set to $0.1$. We used an uncased BERT-base model (\url{https://huggingface.co/bert-base-uncased}) for obtaining encodings of $\{q, q^{cxt}, a\}$.
To obtain encodings of a sequence, two averages were performed: once over all hidden layers, and then over all input tokens.
% Several alternatives were explored, like the CLS token, the second last hidden layer, etc. but they all resulted in significantly worse performance.
Dimension $d = 768$ (from BERT models), and accordingly sizes of the weight matrices were set to %$\boldsymbol{W_1} = 1536 \times 768$ and $\boldsymbol{W_2} = 768 \times 768$.
$\boldsymbol{W_1} = |input| \times 768$ and $\boldsymbol{W_2} = 768 \times 768$, where $|input| = d$ or $|input| = 2d$ (in case we prepend context questions $q^{cxt}_t$).

\myparagraph{Reformulation predictor} % This is used to predict whether follow-up questions are reformulations.
% , during both training and answering times.
To avoid any
% confounding effect and unintended correlations
possibility of leakage from the training to the test data,
this classifier was trained only on the \convquestionsref dev set (as a proxy for an orthogonal source). 
%The sequence classifier at \url{https://bit.ly/2OcpNYw} was used for the fine-tuned BERT sentence pair classification model. 
We fine-tuned the sequence classifier at \url{https://bit.ly/2OcpNYw} for our sentence classification task. 
Positive samples were generated by pairing questions within the same intent, while negative sampling was done by pairing across different intents from the \textit{same conversation}. This ensured lexical overlap in negative samples, necessary for more discriminative learning. 
%The BERT-base-uncased model was used to embed questions.

% \myparagraph{Hardware} 
% %Most
% Experiments used server with 512 GB memory, 6 Intel Xeon CPUs (72 cores total) and 2 NVIDIA Tesla V100 GPUs. 
% % Quadro RTX 8000 graphic card
% %Five additional 8-core Intel Xeon CPUs with 512 GB memory were used as necessary.

\subsection{\conquer configurations}
\label{subsec:config}

The \conquer method has four different setups for training,
%that can be analogously 
that are
evaluated at answering time. These explore two orthogonal sources of noise and stem from 
%a Cartesian product of 
two settings for the user model
and two for the reformulation predictor.

\vspace*{0.1cm}
%\squishlist
%    \item \textbf{User model:} 
\noindent{\bf User model:}\\
    Although \convquestionsref is based on a user study, we do not have continuous access to users during training. 
    Nevertheless, like many other Monte Carlo RL methods, we would like to simulate user behavior for accumulating more interactions that could enrich our training (for example, by performing rollouts). 
    We thus define ideal and noisy user models as follows.
    % In both models, the user reformulates depending only upon the top-1 answer (correct answer at rank 3 when rank 1 is wrong will also trigger a reformulation).
        \begin{itemize}
            \item \textbf{Ideal:} In an ideal user model, we assume that users always behave as expected: reformulating when the generated answer is wrong and only moving on to a new intent when it is correct. Since each intent in the benchmark is allowed up to $5$ times,
            %so when the last reformulation in an intent is reached and the answer is still incorrect, the training proceeds by treating the first question as a follow-up to the fifth one. 
            we loop through the sequence of user utterances within the same intent
            %a conversation 
            if we run out of reformulations. % until a correct answer is obtained, or the desired number of training epochs is reached. 
            \item \textbf{Noisy:} In the noisy variant, the user is free to move on to a new intent even when the response to the last turn was wrong. This  models realistic situations when a user may simply decide to give up on an information need (out of frustration, say). 
            %As a consequence, there are intents with less than five turns despite not having received a correct answer.
            There are at most 4 reformulations per intent in \convquestionsref: so a new information need may be issued after the last one, regardless of having seen a correct response. 
            %which have less than four reformulations. 
            %Such situations are manifested in \convquestionsref when the user moves on to a new information need after the fifth question in an intent: this could either mean that the response was correct, or that the intent was abandoned.
        \end{itemize}
%
%    \item \textbf{Reformulation predictor:}
% \vspace*{0.1cm}
\noindent{\bf Reformulation predictor:}
        \begin{itemize}
            \item \textbf{Ideal:} In an ideal
            %or oracle 
            reformulation predictor, we assume that it is known upfront whether a follow-up question is a reformulation or not (from annotations in \convquestionsref).
            \item  \textbf{Noisy:} In the noisy predictor, we use the BERT-based reformulation detector which 
            may include erroneous predictions.
           % can sometimes make wrong predictions (assuming a reformulation when a new intent is input, and vice versa).
        \end{itemize}
%\squishend

\subsection{Baseline} 
\label{subsec:baseline}

We use the \convex system~\cite{christmann2019look} as the state-of-the-art ConvQA baseline in our experiments. \convex detects answers to conversational utterances over KGs in a two-stage process based on judicious graph expansion: it first detects so-called frontier nodes that define the context at a given turn. Then, it finds high-scoring candidate answers in the vicinity of the frontier nodes. 
%\convex was newly trained on the \convquestionsref train set for fair comparison.
Hyperparameters of \convex were tuned on the \convquestionsref train set for fair comparison.
%\vspace*{-0.1cm}
\subsection{Metrics}
\label{subsec:metrics}

Answers form a ranked list, where the number of correct answers is usually one, but sometimes two or three.
We use three standard metrics for evaluating QA performance: 
i) precision at the first rank (P@1), 
ii) answer presence in the top-5 results (Hit@5) and
iii) mean reciprocal rank (MRR). 
We use the standard metrics of i) precision, ii) recall and iii) F1-score for evaluating context entity detection quality. 
%where we annotated 100 questions with a gold reference set. 
%where each question (in our dev set) has a gold reference set.
%as 
%there is a retrieved and a 
%reference set for these entities for a given question. 
These measures are also used for assessing reformulation prediction performance, where the output is one of two classes: reformulation or not. Gold labels are available from \convquestionsref.

	% !TEX root = ../2021-sigir-fp-conquer.tex
\section{Results and insights}
\label{sec:results}

\subsection{Key findings}
\label{subsec:main-res}

\begin{table*} [t] \small
	%\begin{small}
	% \vspace*{-0.4cm}
	\newcolumntype{G}{>{\columncolor [gray] {0.90}}c}
	\resizebox{\textwidth}{!}{
	\begin{tabular}{l G G G c G G G G G}
		\toprule
		
		\textbf{Method}			&	\textbf{P@1} &	\textbf{Hit@5}	& 	\textbf{MRR}		&	\textbf{RefTriggers}		&			\textbf{Ref = 0}				&	\textbf{Ref = 1}			&	\textbf{Ref = 2}				&	\textbf{Ref = 3}	& \textbf{Ref = 4}					\\ \toprule
		\textbf{\conquer IdealUser-IdealReformulationPredictor}	& 
		$0.339$	&	$0.426$	&	$0.376$	&	$30058$	&	$\boldsymbol{3225}$	&	$292$	&	$154$	&	$70$	&	$56$ \\
		%$0.287$	&	$0.381$	&	$\boldsymbol{0.329}$	&	$32465$	&	$\boldsymbol{2584}$	&	$337$	&	$125$	&	$108$	&	$63$		\\
		\textbf{\conquer IdealUser-NoisyReformulationPredictor}	& 
		%$\boldsymbol{0.293}$	&	$0.379$	&	$\boldsymbol{0.329}$	&	$\boldsymbol{32463}$	&	$2523$	&	$\boldsymbol{404}$	&	$143$	&	$117$	&	$\boldsymbol{90}$  \\
		$0.338$	&	$\boldsymbol{0.429}$	&	$0.377$	&	$30358$	&	$3099$	&	$338$	&	$170$	&	$79$	&	$100$ \\
		\textbf{\conquer NoisyUser-IdealReformulationPredictor}    &
		$\boldsymbol{0.353}$	&	$0.428$	&	$\boldsymbol{0.387}$	&	$\boldsymbol{29889}$	&	$3163$	&	$403$	&	$187$	&	$\boldsymbol{90}$	&	$\boldsymbol{116}$ \\
		%$0.282$	&	$0.382$	&	$0.326$	&	$32657$	&	$2532$	&	$327$	&	$139$	&	$93$	&	$73$\\
		\textbf{\conquer NoisyUser-NoisyReformulationPredictor}	& 
		%$0.281$	&	$\boldsymbol{0.384}$	&	$0.325$	&	32791	&	$2436$	&	$380$	&	$150$	&	$\boldsymbol{121}$	&	$61$ 
		$0.335$	&	$0.417$	&	$0.370$	&	$30726$	&	$2913$	&	$\boldsymbol{425}$	&	$\boldsymbol{216}$	&	$\boldsymbol{90}$	&	$104$ \\ \midrule
		\textbf{\convex}~\cite{christmann2019look}	&	$0.225$	&	$0.257$	&	$0.241$	&	$34861$	&	$1980$	&	$278$	&	$200$	&	$24$	&	$35$\\ \bottomrule
	\end{tabular} }
% 	\\ \raggedright The best value in a column (least for \phrase{RefTiggers}, highest for all others) is in \textbf{bold}. All \conquer variants were statistically signifcantly better than \convex, and showed no differences within themselves ($p < 0.05$). The McNemar's test was performed for binary metrics P@1 and Hit@5, and the 1-tailed $t$-test was performed for real-valued MRR.
	\caption{Main results on answering performance over the \convquestionsref test set.
	%showing \conquer variants and their comparison with the SoTA system \convex. 
	% The best value per column 
	%(least for \phrase{RefTriggers}, highest for all others) 
	% is in \textbf{bold}.
	All \conquer variants outperformed \convex with statistical significance, and required less reformulations than \convex to provide the correct answer.}
	% $p < 0.05$ with McNemar's test for P@1 and Hit@5, and 1-tailed $t$-test for MRR. }
	\label{tab:main-res}
	\vspace*{-0.7cm}
\end{table*}

\begin{table} [t] \small
	% \vspace*{-0.4cm}
	\newcolumntype{G}{>{\columncolor [gray] {0.90}}c}
	\resizebox{\columnwidth}{!}{
	\begin{tabular}{l G c G  c  G}
		\toprule
		\textbf{\conquer/Baseline} &   \textbf{Movies} &\textbf{TV Series} & \textbf{Music}    & \textbf{Books}    & \textbf{Soccer}   \\ \toprule
		\textbf{IdealUser-IdealRef}    	& 
		%$\boldsymbol{0.320}$	&	$\boldsymbol{0.267}$	&	$0.191$	&	$\boldsymbol{0.326}$	&	$\boldsymbol{0.333}$	\\
		$0.320$	&	$0.316$	&	$0.281$	&	$\boldsymbol{0.449}$		&	$\boldsymbol{0.329}$	 \\
		\textbf{IdealUser-NoisyRef}	    & 
		%$0.331$	&	$0.259$	&	$0.222$	&	$0.318$	&	$\boldsymbol{0.333}$	\\
		$0.344$	&	$0.340$	&	$0.303$		&	$0.425$		&	$0.308$ \\
		\textbf{NoisyUser-IdealRef}       
		& $\boldsymbol{0.368}$ &	$\boldsymbol{0.367}$		&	$\boldsymbol{0.324}$ &	$0.413$		&	$\boldsymbol{0.329}$	 \\
		%& $0.309$	&	$0.264$	&	$0.188$	&	$0.322$	&	$0.329$   \\
		\textbf{NoisyUser-NoisyRef}	    & 
		%$0.316$	&	$0.232$	&	$\boldsymbol{0.234}$	&	$0.300$	&	$0.325$	
		$0.327$	&	$0.296$		&	$0.300$	&	$0.381$		&	$0.327$	 \\ \midrule
		\textbf{\convex}~\cite{christmann2019look}  & $0.274$	&	$0.188$	&	$0.195$	&	$0.224$	&	$0.244$	\\ \bottomrule
	\end{tabular}}
%	\\ \raggedright The highest value in a group (metric-domain-system triple) is in \textbf{bold}.
	\caption{P@1 results across topical domains.} % Trends from aggregate table are preserved.} 
	%for \conquer variants on \convquestionsref test set, compared to \convex. 
	% Best values in column in \textbf{bold}.}
	\label{tab:domain-res}
	\vspace*{-1cm}
\end{table}

%MK: maybe sufficient to just add values in text
%\begin{table} [t] \small
%	\centering
	%\resizebox{\columnwidth}{!}{
	% \setlength{\tabcolsep}{0.5em} % for horizontal padding
%	\begin{tabular}{l c  c c}
%		\toprule
%		\textbf{Method}					&	\textbf{P@1}			&	\textbf{Hit@5}				&	\textbf{MRR}		\\	\toprule
%		\textbf{CONQUER NoisyUser-NoisyRef} & $0.259$	& $0.341$  &	$0.293$\\
%		\textbf{CONQUER trained on ConvQuestions} & $\boldsymbol{0.263}$ &	$\boldsymbol{0.343}$ & 	$\boldsymbol{0.298}$\\ \midrule
%	    \textbf{\convex}~\cite{christmann2019look}  & 	$0.184$ & & $0.200$ 
%		\\ \bottomrule
%	\end{tabular} %}
%	\caption{Performance on ConvQuestions.}
%	\label{tab:convquestions}
%	\vspace*{-0.9cm}
%\end{table}

 Tables~\ref{tab:main-res} and~\ref{tab:domain-res} show our main results on the \convquestionsref test set. P@1, Hit@5 and MRR are measured over distinct intents, not utterances. For example, even when an intent is satisfied only at the third reformulation, we deem P@1 $= 1$ (and $0$ when the correct answer is not found after five turns). The effort to arrive at the answer is measured by the number of reformulations per intent (\phrase{RefTriggers}) and by the number of intents satisfied within a given number of reformulations (\phrase{Ref = 1}, \phrase{Ref = 2}, ...). Statistical significance tests are performed wherever applicable: we used the McNemar's test for binary metrics (P@1, Hit@5) and the $t$-test for real-valued ones (MRR, F1). Tests were unpaired when there are unequal numbers of utterances handled in each case due to unequal numbers of reformulations triggered and paired in other standard cases. 1-sided tests
 % were 2-sided when checking for statistical difference among \conquer variants, and 1-sided
 were performed for checking for superiority (for baselines) or inferiority (for ablation analyses). In all cases, null hypotheses were rejected when $p \leq 0.05$. Best values in table columns are marked in \textbf{bold}, wherever applicable.

\myparagraph{\conquer is robust to noisy models} We did not observe any
major differences among the four configurations of \conquer.
% which shows that the RL method can learn good answering models irrespective of the presence of noise in the training data.
Interestingly, some of the noisy versions (\phrase{IdealUser-NoisyRef} and \phrase{NoisyUser-IdealRef}) even achieve the absolute best numbers on the metrics,
%some of the metrics (P@1 and Hit@5, respectively). 
indicating that a certain amount of noise 
and non-deterministic user behavior 
may in fact help the agent to generalize better to unseen conversations.
% (\textit{c.f.} effects of regularization, dropout or masking).
Note that the variants with ideal models are not to be interpreted as potential upper bounds for QA performance: while \phrase{IdealUser} represents a model of systematic user behavior, \phrase{IdealRef} rules out one source of model error. 
% (\phrase{IdealRef} variants are slightly better than their \phrase{NoisyRef} counterparts).

\myparagraph{\conquer outperforms \convex} All variants of \conquer were significantly better than the baseline \convex on the three metrics ($p < 0.05$, 1-sided tests), 
%MK: this is not really the reason why we are better than convex and convex does not learn from qa pairs since it is unsupervised, so maybe we can change the following a bit:
%We attribute this success to the unique ability of \conquer to learn from a sequence of reformulations. \convex, on the other hand, can only learn from
% a subset of
%<question, correct answer> instances in
%the data. Notably, this is not just about \convex, but none of the existing conversational (or otherwise) KG-QA systems~\cite{saha2018complex,guo2018dialog,shen2019multi} can learn from incorrect answers and indirect signals like reformulations.
%MK: proposal:
 showing that \conquer successfully learns from a sequence of reformulations. 
 \convex on the other hand, as well as any of the existing (conversational) 
 %(or otherwise) 
 KG-QA systems~\cite{saha2018complex,guo2018dialog,shen2019multi}, cannot learn from incorrect answers and indirect signals such as reformulations.
 %Table~\ref{tab:convquestions}
 Additionally, \conquer can also be applied in the standard setting were <question, correct answer> instances are available.
 %MK: maybe also mention results with conquer trained on ConvRef but evaluated on Convquestions?
 When trained on the original \convquestions benchmark, that contains gold answers but lacks reformulations, \conquer achieves P@1=$0.263$, Hit@5=$0.343$ and MRR=$0.298$, again outperforming \convex with P@1=$0.184$, Hit@5=$0.219$, MRR=$0.200$. 

\myparagraph{\conquer needs fewer reformulations} %Experimenting with reformulations provides us with the means evaluating interesting facets of the answering process. First, we measure the number of reformulations necessary whenever a particular model satisfactorily answered an information need (\phrase{RefTriggers}).
In Table~\ref{tab:main-res}, \phrase{RefTriggers} shows the number of reformulations needed to arrive at a correct answer (or reaching the maximum of 5 turns).
We observe that \conquer triggers substantially fewer reformulations ($\simeq 30k$) than \convex ($\simeq 34k$).
This confirms an intuitive hypothesis that when a model learns to answer better, it also satisfies an intent faster (less turns needed per intent). Zooming into this statistic (\phrase{Ref = 0, 1, 2, ...}),
we observe that \conquer answers a bulk of the intents without needing any reformulation, a testimony to successful training ($\simeq 3k$ in comparison to $\simeq 2k$ for \convex). The numbers quickly taper off with subsequent turns
% (new intents)
in a conversation, but remain higher than the baseline. \convex relies on a context graph that is iteratively expanded over turns; this often becomes unwieldy at deeper turns.
%MK: I wonder why it is much lower for first question..should we comment on this?
Unlike \convex, we found \conquer's performance to be relatively stable even for higher intent depths (P@1 = %$0.251, 0.304, 0.292, 0.261, 0.328$ 
$0.237$, $0.397$, $0.336$, $0.319$, $0.385$
for intents 1 through 5, respectively).
%\conquer, learning and updating a parameterized policy, is free from such challenges in the symbolic space.

%MK: domain-wise results look a bit different now: books are much higher than rest, TV series not super weak..
\myparagraph{\conquer works well across domains} P@1 results for the five topical domains in \convquestions are shown in Table~\ref{tab:domain-res}. We note that the good performance of \conquer is not just by skewed success on one or two favorable domains (while being significantly better than \convex for each topic), but holds for all five of them (books being slightly better than the others).
% s. Trends from the main results carry over to this per-domain view: only minor differences among \conquer variants, and
% (same statistical significance tests as in Table~\ref{tab:main-res}).
% Like \convex, the quality on TV Series and Music is slightly lower than for the other domains; this is due to the higher number of quantitative and reasoning questions in these domains (e.g., \utterance{How many episodes/albums..?} or \utterance{What was the third album/episode..?}). Such questions do not have entity answers (but aggregates, possibly missing in the KG),
% or require temporal reasoning. 
%Our node-to-node walking agents cannot learn from and answer such questions at this stage.

\subsection{In-depth analysis}
\label{subsec:analysis}

% \begin{table} [t] 
% 	\centering
% %	\columnwidth
% 	%\resizebox*{\columnwidth}{!}{
% 	% \setlength{\tabcolsep}{0.5em} % for horizontal padding
% 	\begin{tabular}{l c  c c}
% 		\toprule
% 		\textbf{Method}		& \textbf{Precision}	    & \textbf{Recall}       & \textbf{F1-Score}     \\	\toprule
% 		All four			& 0.800	                    & 0.539                 & $\boldsymbol{0.612}$  \\  \midrule
% 		- Neighbor overlap	& $\boldsymbol{0.805}$      & 0.525                 & 0.602	                \\
% 		- Lexical match 	& 0.770                     & 0.544                 & 0.605		            \\
% 		- NED		        & 0.116                     & $\boldsymbol{0.630}$  & 0.175                 \\	
% 		- KG prior	    	& $\boldsymbol{0.805}$      & 0.529                 & 0.605                 \\ \bottomrule
% 	\end{tabular} %}
% 	\caption{Ablation on features for context entity detection.}
% %		An asterisk (*) indicates statistical significance of \convex-enabled systems over the strongest baseline in the group.}
% 	\label{tab:startpt}
% 	\vspace*{-0.3cm}
% \end{table}

\begin{table} [t] \small
	\centering
	% \resizebox{\columnwidth}{!}{
	% \setlength{\tabcolsep}{0.5em} % for horizontal padding
	\begin{tabular}{l c c c c c}
		\toprule
		\textbf{All}		& \textbf{No Overlap}	    & \textbf{No Match} & \textbf{No NED} & \textbf{No prior}       \\	\toprule
		%$\boldsymbol{0.612}$ & $0.602$ & $0.605$ & $0.175$* & $0.605$ 
		$\boldsymbol{0.731}$ & $0.726 $ &  $0.728$ &  $0.684$ &  $0.718$  \\ \bottomrule
	\end{tabular} %}
	\caption{Ablation study for context entity detection (F1).}
	\label{tab:startpt}
	\vspace*{-0.7cm}
\end{table}

\begin{table} [t] \small
	\centering
	\resizebox{\columnwidth}{!}{
	\begin{tabular}{l c  c c}
		\toprule
		\textbf{Context model}					            & \textbf{P@1}			    & \textbf{Hit@5}		& \textbf{MRR}		\\	\toprule
		Curr. ques. + Cxt. ent.	
		& $\boldsymbol{0.294}$ & $\boldsymbol{0.407}$ &	$\boldsymbol{0.346}$ 
		%& $\boldsymbol{0.261}$*      & $\boldsymbol{0.382}$*  & $\boldsymbol{0.315}$* 
		\\ \midrule
		Curr. ques. + Cxt. ent. + First ques.  			  
		%& $0.252$	                    & $0.374$	                & $0.305$	                \\
		& $0.254$ &	$0.370$	& $0.305$ \\
		Curr. ques. + Cxt. ent. + First ques. + Prev. ques. 
		%& $0.244$	                    & $0.363$	                & $0.297$	                \\
		& $0.257$ & 	$0.370$	& $0.307$ \\
		Curr. ques. + Cxt. ent. + First refs. + Prev. refs. 
		%& $0.249$	                    & $0.364$	                & $0.300$		        \\
	    & $0.262$ &	$0.382$	& $0.316$ \\	
		\bottomrule
	\end{tabular}}
	\caption{Effect of context models on answering performance.}
	\label{tab:history}
	\vspace*{-0.7cm}
\end{table}

Having shown the across-the-board superiority of \conquer over \convex, we now scrutinize the various components and design choices that make up the proposed architecture. All analysis experiments are reported on the \convquestionsref dev set, and the realistic \phrase{NoisyUser-NoisyRef} \conquer variant is used by default. %, unless otherwise mentioned.  

\myparagraph{All features vital for context entity detection} We first perform an ablation experiment (Table~\ref{tab:startpt}) on the features responsible % for the first block in our pipeline:
for identifying context entities. 
%Since hyperparameters must add up to $1.0$, switching off a particular signal amounts to uniformly redistributing its weights on to the remaining three. 
Observing F1-scores averaged over questions, it is clear that all four factors contribute to accurately identifying context entities (no NED as well as no prior scores resulted in statistically significant drops). It is interesting to understand the trade-off here: a high precision indicates a handful of accurate entities that may not create sufficient scope for the agents to learn meaningful paths. On the other hand, a high recall could admit a lot of context entities from where the correct answer may be reachable but via spurious paths. The F1-score is thus a reliable indicator for the quality of a particular tuple of hyperparameters. 

\myparagraph{Context entities effective in history modeling} After examining features for $E^{cxt}$, let us take a quick look at the effect of $Q^{cxt}$. We tried several variants and show the best three in Table~\ref{tab:history}. While the \conquer architecture is meant to have scope for incorporating various parts of a conversation, we found that explicitly encoding previous questions significantly degraded answering quality (the first row, where $Q^{cxt} = \phi$, works significantly better than all other options).
\phrase{refs.} indicate that embeddings of reformulations for that intent were averaged. Without \phrase{refs.}, only the first question in that intent is used. Results indicate that context entities from the KG 
%MK: context entities are not part of input anymore
%(that are an inseparable part of the policy network's inputs, along with the current question) 
suffice to create a satisfactory representation of the conversation history. Note that these $E^{cxt}_t$ are derived not just from the current turn, but are carried over from previous ones. Nevertheless, we believe that there is scope for using $Q^{cxt}$ better: history modeling for ConvQA is an open research topic~\cite{qu2019bert,qu2019attentive,gupta2021role,qiu2021reinforced} and reformulations introduce new challenges here. % We leave this as promising future work.

\myparagraph{Reformulation predictor works well} A crucial component of the success of \conquer's \phrase{NoisyRef} variants is the reformulation detector. Due to its importance, we explored several options like fine-tuned BERT~\cite{devlin2018bert} and fine-tuned RoBERTa~\cite{liu2019roberta}
% and the Universal Sentence Encoder (USE)~\cite{cer2018universal}
models to perform this classification. RoBERTa produced slightly poorer performance than BERT, which was really effective (Table~\ref{tab:ref-pred}). Prediction of new intents is observed to be slightly easier (higher numbers) due to expected lower levels of lexical overlaps.
% For USE, the cosine similarity between the embeddings of the two questions were performed, and according as this was $> 0.5$ or $\leq 0.5$, we assumed a reformulation or new intent. This examines the effect of using off-the-shelf encodings for this task: however, this resulted in much lower performance (F1 for new intent and reformulations = $0.814, 0.553$ respectively, compared to $0.965, 0.873$ for BERT). This can potentially be improved in practice by tuning the similarity threshold.

\myparagraph{Path label preferable as actions} When defining what constitutes an action for an agent, we have the option of appending the answer entity $e^{ans}$ or the context entity $e^{cxt}$ to the KG path (world knowledge in BERT-like encodings often helps in directly finding good answers). We found, unlike similar applications in KG reasoning~\cite{das2017question,lin2018multi,qiu2020stepwise}, excluding $e^{ans}$ actually worked significantly better for us (Table~\ref{tab:actions}, row 1 vs. 2). This can be attributed to
% our qualifier-enhanced paths in the \conquer KG model, and
the low semantic similarity of answer nodes with the question, that acts as a confounding factor. 
%Excluding $e^{cxt}$ almost has an equivalent performance. % (compare Rows 1 and 3).
%The reason is that an agent selects actions starting at a specific start point ($e^{cxt}$): all of these paths thus have the same start point, resulting in an indistinguishable performance. 
Including $e^{cxt}$ does not change the performance (row 1 vs. 3).
% has almost no effect on the performance.
The reason is that an agent selects actions starting at one specific start point ($e^{cxt}$): all of these paths thus share the embedding for this start point, resulting in an indistinguishable performance. 
The last row corresponds to matching the question to the entire KG fact, which again did not work so well due to the same distracting effect of the answer entity $e^{ans}$.
% : this indicates the importance of correct predicate matching.

\myparagraph{Error analysis points to future work} We analyzed $100$ random samples where \conquer produced a wrong answer (P@1 = 0). We found them to be comprised of: $17\%$ ranking errors (correct answer in top-5 but not at top-1), $23\%$ action selection errors (context entity correct but path wrong), $30\%$ context entity detection errors (including $3\%$ empty set), $23\%$ not in the KG (derived quantitative answers), and $7\%$ wrong gold labels. % , and $3\%$ miscellaneous errors. 
%Anecdotal examples, similar to those in Table~\ref{tab:reformulations} where \conquer retrieves correct answers but not \convex, can be found in our anonymous repository at \url{https://bit.ly/3tHxZjV}.

\myparagraph{Answer ranking robust to minor variations} Our answer ranking (Sec.~\ref{sec:ansgen}) uses cumulative prediction scores (scores from multiple agents added), with P@1 $= 0.294$. We explored variants where we used prediction scores with ties broken by majority voting since an answer is more likely if more agents land on it (P@1 $= 0.291$), majority voting with ties broken with higher prediction scores (P@1 $= 0.273$), and taking the candidate with the highest prediction score without majority voting (P@1 $= 0.294$).
%MK: maybe also say that majority voting as main criterion performed worse (todo: check if significantly worse)
% Majority voting as main criterion works slightly worse, all other
%These 
Most variants were nearly equivalent to each other, showing \textit{robustness of the learnt policy}.

\myparagraph{Runtimes} The policy network of \conquer takes 
%about $9.48$ ms 
about $10.88$ms
to produce an answer, as averaged over test questions in \convquestionsref. 
The maximal answering time was $1.14$s.

\begin{table} [t] \small
	\centering
	\resizebox{\columnwidth}{!}{
	\newcolumntype{G}{>{\columncolor [gray] {0.90}}c}
	\begin{tabular}{l G G G c c c}
		\toprule
		\textbf{Method} & \multicolumn{3}{G}{\textbf{Fine-tuned BERT}}      & \multicolumn{3}{c}{\textbf{Fine-tuned RoBERTa}}   \\ \midrule
		\textbf{Class}  & \textbf{Prec}   & \textbf{Rec}    & \textbf{F1}	& \textbf{Prec} & \textbf{Rec}   & \textbf{F1}		    \\	\toprule
% 		Both            & 0.900     & 0.946	& 0.919	& 0.679 & 0.763 & 0.683               \\
	    New intent		& $0.986$     & $0.944$ & $0.965$	& $0.988$	& $0.924$ & $0.955$ \\
	    Reformulations  & $0.810$     & $0.948$ & $0.873$	& $0.760$ & $0.956$ & $0.847$ \\ \bottomrule
	\end{tabular} }
	\caption{Classification performance for reformulations.}
%		An asterisk (*) indicates statistical significance of \convex-enabled systems over the strongest baseline in the group.}
	\label{tab:ref-pred}
	\vspace*{-0.9cm}
\end{table}

\begin{table} [t] \small
	\centering
	%\resizebox{\columnwidth}{!}{
	% \setlength{\tabcolsep}{0.5em} % for horizontal padding
	\begin{tabular}{l c  c c}
		\toprule
		\textbf{Method}					&	\textbf{P@1}			&	\textbf{Hit@5}				&	\textbf{MRR}		\\	\toprule
		Path	
		%&	$\boldsymbol{0.261}$*	&	$\boldsymbol{0.382}$*	&	$\boldsymbol{0.315}$*			\\ 
			& $\boldsymbol{0.294}$ & $0.407$ &	$\boldsymbol{0.346}$ \\ \midrule
			Path + Answer entity    & $0.275$	& $0.394$ &	$0.329$ \\
		    Context entity + Path  & $0.293$ & $\boldsymbol{0.408}$	& $\boldsymbol{0.346}$ \\
		    Context entity + Path + Answer entity & $0.273$	& $0.398$	& $0.328$
		\\ \bottomrule
	\end{tabular} %}
	\caption{Design choices of actions for the policy network.}
	\label{tab:actions}
	\vspace*{-0.9cm}
\end{table}

\vspace*{-0.1cm}
\section{Related Work}
\label{sec:related}

\myparagraph{QA over KGs} KG-QA has a long history~\cite{yahya2012natural,berant2013semantic,DBLP:conf/rweb/UngerFC14,saha2020question}, 
%starting in about 2012-2013~\cite{yahya2012natural,berant2013semantic}, and
evolving from addressing simple questions via templates~\cite{bast2015more,abujabal17quint} and neural methods~\cite{yih2015semantic,huang2019knowledge}, to more challenging settings of complex~\cite{lu2019answering,bhutani2019learning}, heterogeneous~\cite{oguz2020unified,sun2019pullnet} and conversational QA~\cite{shen2019multi,mueller2019answering}. %The last setting is the one of interest to us, an emerging topic being fuelled by the rise of personal assistants. 
Recent work on ConvQA in particular includes~\cite{saha2018complex,christmann2019look,guo2018dialog,shen2019multi}. 
However, 
%current benchmarks
these methods do not consider question reformulations
in conversational sessions
and solely learn from question-answer training pairs.
%and current systems learn statically from QA pairs, and cannot deal with implicit feedback signals.

\myparagraph{RL in KG reasoning} RL 
%models where agents walk over a knowledge graph, have been
has been 
%successfully applied to the field of 
pursued
for KG reasoning~\cite{das2018go,lin2018multi,xiong2017deeppath,godin2019learning,shen2018m}. Given a relational phrase and two entities, one has to find the best KG path that connects these entities.
%connecting them (or analogous variants). 
This paradigm has been extended to multi-hop QA~\cite{qiu2020stepwise,zhang2018variational}. While \conquer is inspired by some of these settings, the ConvQA
problem is
very different, with multiple entities from where agents could potentially walk and missing entities and relations in the conversational utterances. 

\myparagraph{Reformulations} In the parallel field on text-QA, question reformulation or rewriting has 
%been used to refer to utterances that try to complete ellipses and replace coreferences~
been pursued as conversational question completion
%~\cite{vakulenko2021question,anantha2020open,yu2020few}. 
~\cite{vakulenko2021question,anantha2020open,yu2020few, ren2018conversational, xu2020learning}. 
In our work, we revive the more traditional sense of reformulations~\cite{chang2006query,hassan2013beyond,dang2010query,jansen2009patterns}, where users 
%simply 
pose queries in a different way when system responses are unsatisfactory. 
%Anecdotal analysis of instances in our benchmark \convquestionsref also show proof that reformulations may belong to multiple intuitive categories, beyond just completions. 
%%%GW: this is the related work section, no more side remarks on own work here ...
Several works on search and QA 
apply RL
to automatically generate or retrieve reformulations 
that would proactively
result in the best system response~\cite{buck2018ask,nogueira2017task,das2019multi,ponnusamy2020feedback}. 
%~\cite{xu2020learning,buck2018ask, ren2018conversational}. 
In contrast, \conquer learns from free-form user-generated reformulations. 
Question paraphrases can be considered as proxies of reformulations, without considering system responses.
%(that we tap into, for initializing our baseline RL model) that do not involve system responses. 
Paraphrases have been leveraged in a number of ways in QA~\cite{berant2014semantic,dong2017learning,fader2013paraphrase}. However, such models ignore
%sequence information 
information about sequences of user-system interactions in
% genuine
real conversations.
%that is inherent to the streaming setting that we work in.

\myparagraph{Feedback in QA} Incorporating user feedback in QA is still in its early years \cite{abujabal2018never,kratzwald2019learning,zhang2019interactive, campos2020improving}. 
%only a handful of methods in KG- and text-QA absorb explicit
Existing methods leverage
positive feedback in the form of user annotations to augment the training data. Such explicit feedback is hard to obtain at scale, as it incurs a substantial burden on the user.
In contrast, \conquer is based on the more realistic 
%and challenging 
setting of implicit feedback from reformulations,
which do not intrude at all on the user's natural behavior.
\vspace*{-0.1cm}
\section{Conclusion}
\label{sec:confut}

%In this work, we overcome several conceptual challenges in modeling and evaluation to present the first framework for learning continuously from a stream of utterances and reformulations in conversational question answering over knowledge graphs. 
This work presented
%We demonstrated 
\conquer: an RL-based method for 
conversational QA over KGs,
where users pose ad-hoc follow-up questions in highly colloquial
and incomplete form.
For this ConvQA setting, \conquer is the first method that leverages implicit negative feedback when users reformulate
previously failed questions.
%
%that answers ad hoc and incomplete utterances in ConvQA without explicit structured queries, where agents walk from context entities to candidate answers over KG paths using predictions from a policy network. 
%%%GW: no need for this detail in conclusion, readers have read the paper already
%\conquer was shown to be robust to noisy user models and reformulation predictors, and all variants significantly outperformed the strong baseline system \convex. 
Experiments with a benchmark based on a user study showed that
\conquer outperforms the state-of-the-art ConvQA baseline \convex \cite{christmann2019look}, and
that \conquer is robust to various kinds of noise.
%
%%%GW: now, one very short sentence on future work
%However, this is only the first step towards a new QA setting for learning without static QA pairs, and opens up several interesting directions of future work. Some of the more promising avenues would be better modeling of the conversational context by judiciously introducing candidate answers, exploring varying levels of noise in user behavior, and applying similar frameworks to other settings like multi-hop QA.
% framework applicable beyond QA
%%%GW: or nothing on outlook

\myparagraph{Acknowledgments} We would like to thank Philipp Christmann from MPI-Inf for helping us with the experiments with CONVEX.

% \clearpage

	%\clearpage
	\bibliographystyle{ACM-Reference-Format}
	\bibliography{conquer}
	
\end{document}